\documentclass[preprint,3p,twocolumn]{elsarticle}
\usepackage{geometry}
\usepackage{fancyhdr}

\biboptions{numbers,sort&compress}
\usepackage[resetlabels]{multibib}

\journal{}

\biboptions{comma,square,sort&compress}

\usepackage{amsmath}
\usepackage{amstext}
\usepackage{amssymb}
\usepackage{amsfonts}
\usepackage{bm} 
\usepackage[colorlinks,linkcolor=black, citecolor=black, urlcolor=blue, plainpages=false, pdfpagelabels]{hyperref}

\usepackage[separate-uncertainty=true, output-decimal-marker={.}, per-mode=symbol, parse-numbers=true]{siunitx}
\DeclareSIUnit{\wtpercent}{wt\percent}
\DeclareSIUnit{\atpercent}{at.\percent}
\DeclareSIUnit{\dnnt}{R_s/V_s}
\DeclareSIUnit{\dnnv}{V_s}
\DeclareSIUnit{\dnnr}{R_s}

\begin{document}
\begin{frontmatter}

\title{Convective effects on columnar dendritic solidification -- \\ A multiscale dendritic needle network study }

\author{T. Isensee$^{a,b}$}
\author{D. Tourret$^{a,*}$}
\address{$^a$ IMDEA Materials Institute, C/ Eric Kandel 2, 28906, Getafe, Madrid, Spain.}
\address{$^b$ Department of Materials Science, Polytechnic University of Madrid/Universidad Polit\'ecnica de Madrid, E.T.S. de Ingenieros de Caminos, 28040, Madrid, Spain.}

\cortext[cor1]{damien.tourret@imdea.org}

\begin{abstract}

Gravity-induced buoyancy, inevitable in most solidification processes, substantially alters the dynamics of crystal growth, such that incorporating fluid flow in solidification models is crucial to understand and predict key aspects of microstructure selection. 
Here, we present a multi-scale Dendritic Needle Network (DNN) model for directional solidification that includes buoyant flow in the liquid, and apply it to a range of alloys and growth conditions. 
After a brief presentation of the model, we study the selection of stable primary dendrite arm spacings in Al-$\SI{4}{\atpercent}$-Cu and in Ti-$\SI{45}{\atpercent}$-Al alloys under different gravity levels, comparing both applications to published phase-field results and experimental measurements. 
Then, we simulate the oscillatory growth behavior recently reported via X-ray in situ imaging of directional solidification of nickel-based superalloy CMSX-4.
In this last application, the DNN simulations manage to reproduce the oscillatory growth behavior, and hence permit identifying the fundamental mechanisms behind the oscillatory growth regime.
In particular, we show that sustained oscillations occur when the average liquid flow velocity is close to the crystal growth velocity, and that primary dendritic spacings also play a crucial role in the oscillatory behavior.

\end{abstract}


\end{frontmatter}

 \thispagestyle{fancy} 


\section{Introduction}
\label{sec:intro}

Solidified metals and alloys predominantly exhibit dendritic microstructures with geometrical features that have a direct effect on the thermo-mechanical properties of materials \cite{trivedi1994}. The primary dendritic arm spacing, for instance, may determine to a large extent the ultimate tensile strength \cite{quaresma2000,osorio2002}. Thus, especially for materials exposed to high temperatures and stresses, it is of tremendous importance to predict and control such characteristic length scales emerging during solidification.

Dendritic morphologies result from a complex interplay between phenomena on different length scales: from capillarity effects at the atomistic scale of the solid-liquid interface to macroscopic heat and solute transport in the liquid \cite{langer1980,trivedi1994}. Within the past decades, many theoretical approaches have addressed the selection of dendritic patterns at different length scales, primarily focusing on the fundamental phenomena of capillarity and diffusion \cite{ivantsov1947,barbieri1989,benamar1993}.
However, convective transport in the liquid phase was also reported to have a great influence on dendritic microstructure selection \cite{mehrabian1970,nguyen1989}.

Buoyant flow in the liquid phase is primarily due to the gradients in temperature and solute concentration resulting from crystal growth, combined with the effect of gravity. 
Experiments in microgravity have been carried out in order to circumvent the effect of gravity-induced buoyancy \cite{glicksmann1994,nguyen2005,nguyen2017}. 
However, melt flow is inevitable under realistic Earth-based experimental and industrial conditions.
Fluid flow adds an extra level of complexity to the relatively well-studied dendritic growth under purely diffusive conditions, but its fundamental understanding remains both paramount and challenging. 

The consequences of fluid flow on the resulting dendritic microstructures are multiple.
The stirring of the liquid phase contributes to a reduction of the solute boundary layer ahead of the growing front, which may extend the range of stable velocities for a planar solid-liquid interface \cite{clarke2017microstructure}.
Fluid flow also substantially affects the selection of microstructural length scales, such as primary dendritic spacings \cite{dupoy1992,bataile1994, trivedi2002effect}.
In spite of these observations, the understanding of fundamental relationships between processing and microstructures during solidification in the presence of convection remains incomplete.
A reason for this knowledge gap is the lack of modeling approaches for quantitative simulations at the relevant length/time scales. 
This article uses a recently proposed multiscale model to address some of these outstanding gaps.

Convective effects on directional solidification (DS) are particularly important in the context of Nickel (Ni) superalloys for aeronautical applications.
Indeed, single-crystal turbine blades are typically produced by DS \cite{pollock2006}, and undesirable defects, such as freckles, are closely tied to convective transport of solute species in the liquid \cite{giamei1970nature, pollock1996breakdown, auburtin2000freckle}. 
In this context, the emergence of in situ imaging techniques for metallic alloys, e.g. the use of time-resolved X-ray radiography, has allowed a substantial advance in the study of gravity-induced flow and its consequences on microstructure selection \cite{bogno2011,shevchenko2013,clarke2015x,reinhart2020}.
Among such recent observations, an oscillatory growth regime was observed during DS of CMSX-4, a Ni-based superalloy commonly used for single-crystal turbine blades \cite{reinhart2020}. 
This unstable growth of the solidification front has been linked to the presence of buoyant flow in the liquid, but the fundamental mechanisms behind these oscillations remain to be explored and explained on a quantitative basis, due to the lack of adequate modeling method~\cite{reinhart2020}.
In this article, we reproduce the oscillatory growth during DS of CMSX-4 and bring quantitative clarifications on this nontrivial behavior.

In terms of modeling, the phase-field (PF) method, implicitly tracking the solid-liquid interface, has for decades been the computational method of choice to simulate dendritic growth \cite{boettinger2002phase}. 
Integrating melt flow within PF models has allowed the study of dendrite morphologies under forced \cite{beckermann1999,jeong2001,jeong2003} and natural convection \cite{steinbach2009}, and the exploration of the effect of fluid flow on primary spacing in columnar dendritic arrays \cite{steinbach2009,viardin2020a,viardin2020b}. 
However, simulation domains have for the most part remained limited in size to a handful of primary dendrites.
Recent numerical methods have enabled substantial acceleration, e.g. using parallelization on Graphics Processing Units (GPUs) and/or using the Lattice Boltzmann method \cite{takaki2015,takaki2017,takaki2018}. Still, due to the scale separation between dendritic tips and solute transport, simulations of dendritic growth with fluid flow at experimentally relevant length and time scales with the PF method remain challenging, unless using advanced algorithms and tremendous computational resources \cite{sakane2020two}.

In order to address these computational limitations, the multiscale Dendritic Needle Network (DNN) model \cite{tourret2013a,tourret2016} was designed to bridge the scale gap between PF and coarse-grained models. The dendritic structure is described by a hierarchical network of thin parabolic-shaped needles. It was shown to be well suited for modeling spacing selection in binary alloys \cite{bellon2021}. For equiaxed growth, the model was extended to include liquid melt flow in two dimensions (2D) \cite{tourret2019} and recently three dimensions (3D) \cite{isensee2020}.

In this article, we present a 2D formulation of the DNN model applied to directional solidification conditions, including convective transport in the melt (Sec.\,\ref{sec:model}).
We verify the predictions of the DNN model in terms of primary dendrite arm spacings, by comparing them to results of independent PF simulations and experimental data for aluminum-copper \cite{steinbach2009} (Sec.\,\ref{subsec:verification_al-cu}) and titanium-aluminum \cite{viardin2020a,viardin2020b} (Sec.\,\ref{subsec:verification_ti-al}) alloys.
Finally (Sec.\,\ref{sec:nickel}), we simulate the buoyancy-induced oscillatory growth observed in CMSX-4 directional solidification \cite{reinhart2020}, which enables a deeper exploration of its key underlying mechanisms.

\section{Model}
\label{sec:model}

The model used here, and its numerical implementation, are direct extensions of our previous works.  
Therefore, we only provide a brief introduction to the key concepts and equations of the method, while all further technical details can be found in earlier articles \cite{tourret2016,tourret2019}.

\subsection{Sharp-interface model}

We consider a binary alloy of nominal solute concentration $c_\infty$ in the dilute limit where the interface solute partition coefficient $k=c_s/c_l$ between equilibrium concentrations of solid ($c_s$) and liquid ($c_l$) phases can be considered constant.
The temperature field is assumed to follow the frozen temperature approximation $T=T_0+G(x-V_p t)$, with a reference temperature $T_0$, a constant temperature gradient $G$ and a pulling velocity $V_p$. 
Here, the reference temperature $T_0$ is chosen as the alloy solidus temperature $T_s$ at its nominal concentration $c_\infty$.
For moderate growth velocities, kinetic undercooling can be neglected, such that the equilibrium of the solid-liquid interface can be written via the Gibbs-Thomson relation \cite{tourret2016}
\begin{equation}
\label{eq:gibbs_thomson}
 \frac{c_l}{c_l^0} = 1 - (1-k)d_0 f(\theta)\kappa - (1-k)\frac{x-V_p t}{l_T},
\end{equation}
where $c^0_l = (T_M-T_L)/|m|=c_\infty/k$ is the liquid equilibrium concentration of a flat interface at $T_0$, $d_0 = \Gamma/\left[|m|(1-k)c_l^0\right]$ is the capillary length at $T_0$ with $\Gamma$ the interface Gibbs-Thomson coefficient, $f(\theta)$ expresses the dependence of the interface stiffness upon its orientation ($\theta$), $\kappa$ is the interface curvature, and the thermal length $l_T = |m|(1-k)c^0_l/G$ corresponds to the freezing range of the alloy. 
The Gibbs-Thomson equation \eqref{eq:gibbs_thomson} is combined with a statement of solute conservation at the solid-liquid interface that takes the form of the Stefan condition 
\begin{equation}
\label{eq:stefan}
(1-k)c_l \bm{v}_n = D\nabla c|_i ,
\end{equation}
where $\bm{v}_n$ is the interface velocity, $D$ is the solute diffusion coefficient in the liquid phase, assuming that diffusion in the solid is negligible, and $\nabla c|_i$ denotes the solute concentration gradient in the liquid at the interface.
Finally, the sharp-interface problem is completed by an equation for the transport of solute in the bulk, which may, in the vicinity of the interface, be considered to follow the diffusion equation 
\begin{equation}
\label{eq:diffu}
\partial_t c = D\nabla^2 c,
\end{equation}
but may also incorporate additional (e.g. advective) terms in the bulk liquid further from the interface (See Sec.\,\ref{sec:navierstokes}).

\subsection{Reduced solute field}

Introducing the reduced solute field $U = (c^0_l-c)/[(1-k)c^0_l]$, the Gibbs-Thomson relation \eqref{eq:gibbs_thomson}, i.e., the interface equilibrium concentration can be written as
\begin{equation}
\label{eq:gibbs_thomson_U}
 U_i = d_0 f(\theta)\kappa + \frac{x-V_p t}{l_T}.
\end{equation}
with the far-field condition $U_i(x\rightarrow+\infty) = 1$. 
The diffusion equation and Stefan condition for the non-dimensional field $U$ hence become

\begin{align}
\label{eq:diffusion_U}
 \partial_t U &= D\nabla^2 U,\\
 \label{eq:stefan_U}
 \left[1-(1-k)U_i\right]\bm{v}_n &= D\partial_n U|_i.
\end{align}

\subsection{Solvability condition}

Several studies \cite{barbieri1989, benamar1993} have shown that at the small scale of the dendritic tip radius $R$, the free boundary problem defined by \eqref{eq:gibbs_thomson_U}-\eqref{eq:stefan_U} only has a solution if the microscopic solvability condition holds, which reads
\begin{equation}
\label{eq:r2v}
 R^2 V = \frac{2Dd^*_0}{\sigma} = \frac{1}{1-(1-k)U_t}\frac{2Dd_0}{\sigma},
\end{equation}
with $d^*_0$ the capillary length expressed at the tip temperature, $\sigma$ the tip selection parameter, and $U_t = (x_t-V_p t)/l_T$ the equilibrium concentration at the tip position $x_t$, neglecting curvature and kinetic undercooling.

\subsection{Flux intensity factor}

At a scale much larger than the tip radius $R$, where the curvature of a needle is negligible, but much smaller than the diffusion length $l_D$, we can integrate the Stefan condition \eqref{eq:stefan_U} along a parabolic tip \cite{tourret2016}, leading to 
\begin{equation}
\label{eq:rv2}
 RV^2 = \frac{2D^2\mathcal{F}^2}{\left[1-(1-k)U_t\right]^2 d_0}.
\end{equation}
The flux intensity factor (FIF) $\mathcal{F}$ measures the normal solute flux towards the dendrite along the contour $\Gamma_0$ along the interface up to a distance $a$ behind the tip. It is defined as
\begin{equation}
\mathcal{F} := \frac{1}{4\sqrt{a/d_0}}\int_{\Gamma_0} (\partial_n U)\,dS,
\end{equation}
where $\partial_n U$ is the flux normal to the interface.
In practice, one can choose a more convenient integration domain $\Gamma_i$ (here circular), that encloses the area $\Sigma_i$ around the needle tip \cite{tourret2019,isensee2020}. 
Using the divergence theorem and assuming a Laplacian solute field in the domain moving with velocity $V$, the integral of the FIF can be calculated by
\begin{equation}
\label{eq:fif}
  4\mathcal{F}\sqrt{a/d_0} = \int_{\Gamma_i} \left(\partial_{n^*} U\right) dS +\frac{V}{D}\int_{\Sigma_i}\left(\partial_x U\right) dA,
\end{equation}
with $\partial_{n^*} U$ the flux across the $\Gamma_i$ integration contour using an outwards pointing normal vector $\bm{n^*}$, for a needle growing in the $x$-direction \cite{tourret2019,isensee2020}.

\subsection{Solute transport}
\label{sec:navierstokes}
On the large scale of the diffusion length $l_D=D/V$ and above, the dendrites appear as thin needles, their curvature can be neglected, and the Gibbs-Thomson relation \eqref{eq:gibbs_thomson_U} can be approximated by
\begin{equation}
\label{eq:equilU}
 U_i = \frac{x-V_p t}{l_T}.
\end{equation}
At all times, Eq.~\eqref{eq:equilU} is imposed as an internal boundary condition over the entire needle network, as it represents the fact that the solid-liquid interface is at equilibrium.

In the bulk liquid we consider solute transport by not only diffusion but also by (buoyancy-driven) convection, by solving the incompressible Navier-Stokes equation 
\begin{equation}
\label{eq:ns}
 \rho\left[\partial_t\bm{v}+(\bm{v}\cdot\nabla)\bm{v}\right] = \bm{F} - \nabla p + \eta\nabla^2\bm{v}
\end{equation}
for the fluid velocity $\bm{v}$, where $\rho$ is the fluid density, $p$ is the pressure, $\eta$ is the viscosity and $\bm{F}$ represents external forces. The incompressibility condition reads
\begin{equation}
\label{eq:poisson}
 \nabla\cdot\bm{v}=0.
\end{equation}
Here, we only account for external buoyancy forces due to solute concentration gradients, considering that they are typically dominant over those induced by temperature gradients.
Hence, we use the Boussinesq approximation for the buoyancy force term,
\begin{equation}
 \bm{F} = \rho^l_\infty\bm{g}\left[1-\beta_c(c-c_\infty)\right],
\end{equation}
with a solutal expansion coefficient
\begin{equation}
\label{eq:solutal_expansion_coefficient}
 \beta_c = -\frac{1}{\rho^l_\infty}\frac{\partial\rho}{\partial c}\Bigr|_{c=c_\infty},
\end{equation}
evaluated at the nominal concentration, where the liquid density is $\rho^l_\infty$.
The transport of solute in the liquid with fluid velocity $\bm{v}$ is thus described by the advection-diffusion equation
\begin{equation}
\label{eq:U}
 \partial_t U + \nabla\cdot(\bm{v}U) = D\nabla^2 U.
\end{equation}

\subsection{Implementation}

The resulting model consists in solving the incompressible Navier-Stokes problem \eqref{eq:ns}-\eqref{eq:poisson} and the advection-diffusion equation \eqref{eq:U} in the liquid phase.
An equilibrium condition on the concentration field, Eq.~\eqref{eq:equilU}, and a null velocity are imposed over a network of parabolic branches.
At each time, the tip radius and growth velocity of each individual branch is calculated from Eqs~\eqref{eq:r2v}-\eqref{eq:rv2}, where the FIF is integrated according to Eq.~\eqref{eq:fif}.
The numerical resolution of the model and its implementation are presented in detail in \cite{tourret2019}.
Essentially, using a finite difference spatial discretization on a staggered grid, the Navier-Stokes equations are solved using a projection method \cite{chorin1968} and an iterative successive over-relaxation method \cite{frankel1950, young1954} is used for the incompressibility condition. The time-stepping is carried out with an explicit Euler method. The code is implemented in the C-based CUDA programming language for Nvidia GPUs, which allows a substantial acceleration via parallelization.

\section{Gravity effect on primary spacing selection}
\label{sec:verification}

\begin{table*}[ht!]
\centering
\begin{tabular}{lccc}
\hline
Property & Symbol & Value & Unit\\
\hline
Nominal composition & $c_\infty$ & $4$ & $\si{\atpercent}$ \\ 
Liquidus slope & $m$ & $-1.6$ & $\si{\kelvin\per\atpercent}$ \\
Partition coefficient & $k$ & $0.14$ & \\
Liquid diffusivity & $D_l$ & $\num{3e-9}$ & $\si{\meter^2\per\second}$ \\
Kinematic viscosity & $\nu$ & $\num{5.7e-7}$ & $\si{\meter^2\per\second}$ \\
Solutal expansion coefficient & $\beta_c$ & $\num{-e-2}$ & $\si{\per\atpercent}$ \\
Interfacial energy anisotropy (PF) & $\epsilon$ & $\num{2e-2}$ & \\
Tip selection parameter (DNN) & $\sigma$ & \num{0.153} & \\
\hline
Temperature gradient & $G$ & $\num{e4}$ & $\si{\kelvin\per\meter}$ \\
Pulling velocity & $V_p$ & $\num{4e-5}$ & $\si{\meter\per\second}$ \\
\hline
Finite difference grid spacing & $h$ & $\numrange{0.8}{1.25}$   & $R_s$ \\
FIF integration radius & $r_i$ & $4$ &$h$ \\
Parabola truncation radius & $r_\text{max}$ & $\num{1}$   & $r_i$ \\
Upwind parameter & $\omega_\text{up}$ & $0.9$ & \\
Successive Over Relaxation parameter  & $\omega_\text{SOR}$ & $1.1$ & \\
SOR residual required for convergence & $\overline{r}_\text{SOR}$ & $\num{e-3}$ & \\
Time step safety factor & $K_{\Delta t}$ & $\numrange{0.3}{0.6}$ & \\
\hline
\end{tabular}
\caption{Material and processing parameters for directional solidification of Al-$\SI{4}{\atpercent}$-Cu from \cite{steinbach2009} and numerical parameters (see ref. \cite{tourret2019} for details).\label{tab:al_cu_0}}
\end{table*}

For a given alloy under given processing conditions, the primary dendritic spacing, $\lambda_1$, is known to be selected within a broad range \cite{han1994primary, hunt1996numerical, echebarria2010onset, bellon2021}.
Below a minimum spacing $\lambda_\text{min}$, dendrites get eliminated through solute interaction with neighbors.
Above a maximum spacing $\lambda_\text{max}$, dendritic side-branching occurs and new primary branches emerge.
Moreover, the solute transport regime is well acknowledged to greatly influence spacing selection \cite{dupoy1992, bataile1994, trivedi2002effect}.

In the first two applications of the DNN model, we study the selection of primary dendritic spacing under different gravity conditions. 
To do so, we consider two independent studies for Al-Cu \cite{steinbach2009} and Ti-Al \cite{viardin2020b} alloys.
Both studies rely on 2D phase-field simulations using a multi-phase field approach coupled to a Navier-Stokes solver, hence providing a fair quantitative comparison with our 2D DNN simulations results.
These quantitative comparisons constitute a sound verification -- against the reference PF results -- and validation -- against the corresponding experimental data -- of the DNN method.

\subsection{Spacing selection in Al-Cu alloy}
\label{subsec:verification_al-cu}

Primary spacing selection via elimination ($\lambda_\text{min}$) in  directional solidification in a buoyancy-driven flow was addressed with the PF method for Al-$\SI{4}{\atpercent}$-Cu \cite{steinbach2009}. 
There, the effect of gravity strength was investigated, and the following scaling law was proposed 

\begin{align}
\label{eq:steinbach1}
 a_0 g &= \lambda^{-7} - \lambda^{-4}\quad&\text{for }g\geq 0\\
\label{eq:steinbach2}
 a_0 g &= -\lambda^2 + \lambda^{-4}\quad&\text{for }g\leq 0
\end{align}
which describes the ratio $\lambda=\lambda_1/\lambda_0$ between the primary dendritic spacing $\lambda_1$ and its value in absence of gravitational forces $\lambda_0$, when gravity and growth are in the same direction ($g>0$) or in opposite directions ($g<0$). A prefactor value $a_0=5$ was found to yield a good agreement to PF results \cite{steinbach2009} and experimental measurements \cite{bataile1994}.

\subsubsection{DNN simulations}

We carried out DNN simulations of directional solidification using similar alloy and processing parameters as in Ref.~\cite{steinbach2009}.
Thermophysical alloy properties, processing conditions, and numerical parameters (see detailed definitions in Ref.~\cite{tourret2019}) are listed in Table \ref{tab:al_cu_0}.
 
Instead of reduced-size PF simulations \cite{steinbach2009}, DNN simulations are performed over entire dendritic arrays of at least 13 (and up to 51) primary dendrites growing together at steady state.
The simulations are initialized with several parallel and evenly spaced needles with their tips located at the liquidus temperature. The envelope joining all tips is meant to approximate a planar front. The initial solute distribution is given by $U(x<l_T, y) = x/l_T$ and $U(x>l_T, y) = 1$. The simulations are carried out on a moving domain, meaning that the most advanced needle tip stays at a fixed position within the computational domain. 

The boundary conditions are periodic in the $y$-direction (laterally).
On the top and bottom boundaries (normal to the growth direction $x$), we set free-slip conditions with $v_x = 0$ for the fluid flow, meaning that flow through the boundary is not allowed. The diffusion field on the top boundary is set to a constant value of $U=1$, which corresponds to the nominal concentration $c_\infty=\SI{4}{\atpercent}$Cu. On the bottom boundary, we set no-flux (mirror) conditions with $\partial U/\partial x = 0$. 

The finite difference grid spacing, $h$, is set between $\SI{4.6}{\micro\meter}$ (for $g=3g_0=\SI{29.43}{\meter\per\second^2}$) and  $\SI{7.1}{\micro\meter}$ (for all other $g$), which corresponds to $\num{0.8}\leq h/R_s\leq\num{1.25}$, with $R_s = \SI{5.7}{\micro\meter}$ the theoretical steady state tip radius for $g=0$ \cite{tourret2016,tourret2019}.
The contour used to integrate the flux intensity factor is a circle centered on the tip with a radius $r_i=4h$ and the parabolic tips are bound to a maximum radius  $r_\text{max}=r_i$ \cite{tourret2016,tourret2019}.

The most advanced dendrite tip is fixed at a height of $\SI{0.9}{\milli\meter}$, and the domain is initialized with between 14 and 118 evenly spaced parallel primary dendrites.
Due to the competition for solute among the dendrites, individual dendrites progressively get eliminated, i.e., they leave the moving domain.
Eventually, a growth state with stable primary dendrite arm spacing is reached when no more elimination events occur.
We determined the stability range of primary dendritic spacings from several simulations, varying domain sizes, initial needle distributions (and hence initial $\lambda_1$), and gravity acceleration (direction and strength). 
From the initial and final distributions of primary dendrites, we extract the maximum unstable spacing and the minimum stable spacing.
They provide an estimate of the range in which the minimum spacing with respect to elimination, $\lambda_\text{min}$, is expected.

Simulations were performed with different domain sizes (250\,000 to 600\,000 grid points) and simulated times (300 to 1\,500\,s).
Using a single Nvidia RTX 2080Ti GPU, computation times for $g\neq0$ ranged from $\num{3}$ to $\num{16}$ days, while simulations at $g=0$ lasted just a few hours. 

\subsubsection{Results and discussion}
\label{sec:alcu_resu}

\begin{figure}[t!]
\centering
\includegraphics[width=\columnwidth]{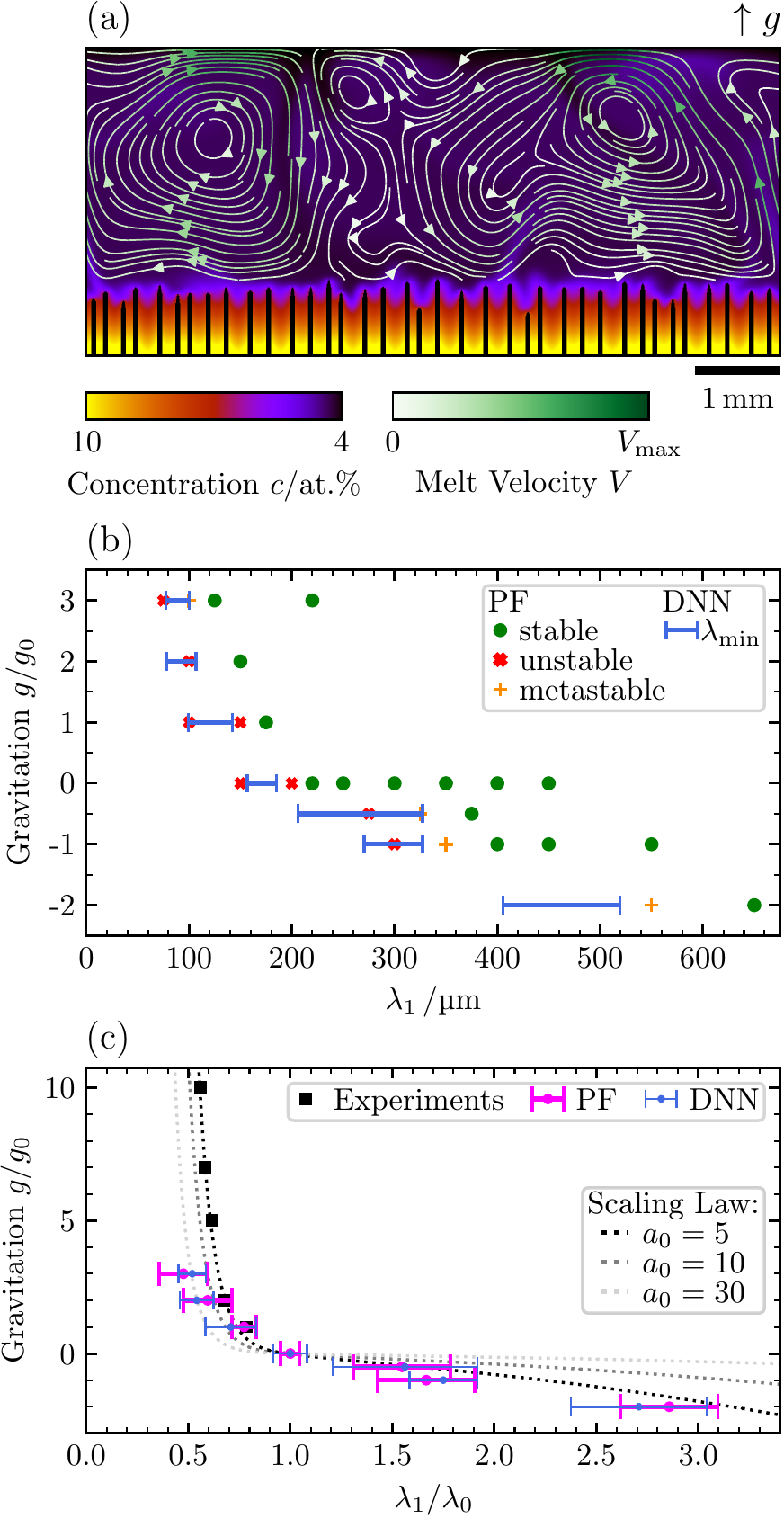}
\caption{(a) Final state at $t=\SI{300}{\second}$ of a DNN simulation with gravity $g=g_0=\SI{9.81}{\meter\per\second^2}$ pointing upwards, i.e. in the growth direction. 
Streamlines show the fluid flow above the solid region, with $V_\text{max}=55\,V_p=\SI{2.2e-3}{\meter\per\second}$. 
(b) Stable and unstable spacing distribution in Al-$\SI{4}{\atpercent}$-Cu, extracted from PF \cite{steinbach2009} and DNN (this work) simulations for different levels of gravity with $g>0$ pointing in growth direction and $g<0$ pointing against growth direction. (c) Scaling laws \eqref{eq:steinbach1}-\eqref{eq:steinbach2} compared to experimental data \cite{bataile1994} and simulations for different values of the prefactor $a_0$. \label{fig:al_cu_combined_0}}
\end{figure}

Fig.\,\hyperref[fig:al_cu_combined_0]{1a} shows the final state at $t=\SI{300}{\second}$ of a simulation with $g=\SI{9.81}{\meter\per\second^2}$ pointing in the growth direction of the dendrites. 
The domain has $N_x\times N_y = 1150\times 510$ grid points, which makes it $L_y=\SI{8.2}{\milli\meter}$ wide and $L_x=\SI{3.6}{\milli\meter}$ high.
Of the initially placed $115$ dendrites only $37$ remain after growth competition and elimination.
In this simulation, as in several others, the solute flow contributes to the stabilization of some dendrites slightly trailing behind the leading ones, but eventually growing at a velocity $V_p$ without being eliminated. 
The presence of these metastable spacings are consistent with PF results \cite{steinbach2009}.

Fig.\,\hyperref[fig:al_cu_combined_0]{1b} shows the stable and unstable spacing distribution for each investigated gravity level, in comparison with PF results from Ref.~\cite{steinbach2009}. 
The minimum spacings $\lambda_\text{min}$ predicted by the DNN exhibits a good agreement with the PF predictions.
At $g\geq0$, the DNN-predicted spacings are slightly smaller, but the discrepancy on average $\lambda_\text{min}$ values remains within about 26\% between DNN and PF results. 
This discrepancy may stem from the fact that PF simulations used a limited domain size (e.g. domain height of $\SI{600}{\micro\meter}$ corresponding to $\SIrange{10}{25}{\percent}$ of the current simulations), commensurate with computational capabilities at the time.

The scaling laws derived in Ref.~\cite{steinbach2009} for upwards downwards gravity directions, i.e. Eqs~\eqref{eq:steinbach1}-\eqref{eq:steinbach2}, are compared to PF, DNN, and experimental \cite{bataile1994} results in Fig.\,\hyperref[fig:al_cu_combined_0]{1c}. 
The spacings for $g\leq0$, in good agreement with PF results, are also in good agreement with the scaling law with the prefactor $a_0=5$ identified in Ref.~\cite{steinbach2009}.
For $g\geq0$, our results still follow the expected trend, but the prefactor seems closer to $a_0\approx30$, but since this value severely overestimates spacings at $g\leq0$, $a_0\approx5$ remains a nearly optimal value.

\begin{table*}[ht!]
\centering
\begin{tabular}{lccc}
\hline
Property & Symbol & Value & Unit\\
\hline
Nominal composition & $c_\infty$ & $45$ & $\si{\atpercent}$ \\ 
Liquidus slope & $m$ & $-11.26$ & $\si{\kelvin\per\atpercent}$ \\
Partition coefficient & $k$ & $0.9$ & \\
Liquid Diffusivity & $D_l$ & $\num{3e-9}$ & $\si{\meter^2\per\second}$ \\
Gibbs-Thomson coefficient & $\Gamma$ & $\num{1.61e-7}$ & $\si{\kelvin\meter}$ \\
Kinematic viscosity & $\nu$ & $\num{1.89e-6}$ & $\si{\meter^2\per\second}$ \\
Solutal expansion coefficient & $\beta_c$ & $\num{4.784e-3}$ & $\si{\per\atpercent}$ \\
Interfacial energy anisotropy (PF) & $\epsilon$ & $\num{1.1e-2}$ & \\
Tip selection parameter (DNN) & $\sigma$ & \num{0.145} & \\
\hline
Temperature gradient & $G$ & $\num{1.2e4}$ & $\si{\kelvin\per\meter}$ \\
Pulling velocity & $V_p$ & $\num{2.5e-5}$ & $\si{\meter\per\second}$ \\
\hline
Finite difference grid spacing & $h$ & $\num{1.75}$ & $R_s$ \\
FIF integration radius & $r_i$ & $\num{7}$   &$R_s$ \\
Parabola truncation radius & $r_\text{max}$ & $\num{7}$ & $R_s$ \\
Upwind parameter & $\omega_\text{up}$ & $0.9$ & \\
Successive Over Relaxation parameter  & $\omega_\text{SOR}$ & $1.1$ & \\
SOR residual required for convergence & $\overline{r}_\text{SOR}$ & $\num{e-3}$ & \\
Time step safety factor & $K_{\Delta t}$ & $\numrange{0.2}{0.6}$ & \\
\hline
\end{tabular}
\caption{Material and processing parameters for directional solidification of Ti-$\SI{45}{\atpercent}$-Al from \cite{viardin2020b} and numerical parameters (see ref. \cite{tourret2019} for details).\label{tab:ti_al_0}}
\end{table*}

Experimental measurements \cite{bataile1994}, only available for $g\geq g_0$, are close to the higher values of $\lambda_\text{min}$ assessed by both PF \cite{steinbach2009} and current DNN results. 
This small discrepancy between experiments and simulations may be attributed to uncertainties in alloy parameters, but also importantly to dimensionality -- comparing 2D simulations with 3D experiments. 
This effect is not trivial. 
Indeed, on the one hand, two-dimensional simulations are known to enhance diffusive interaction among dendrites, consequently overestimating 3D spacings even in diffusive conditions \cite{tourret2015}.
Yet, on the other hand, convection is expected to reduce the length of diffusive interaction, and hence reduce the spacing.
This latter effect is enhanced even further by the fact that fluid velocities, and their consequences on crystal growth, may also be severely overestimated in 2D simulations \cite{jeong2001,isensee2020}.
The current results, from both DNN and PF methods, suggest that the second effect, reducing spacings in 2D simulations, may be dominant.

\subsection{Spacing selection in Ti-Al alloy}
\label{subsec:verification_ti-al}

While the previous section was focused on instabilities in columnar growth due to the elimination of dendrites (when $\lambda<\lambda_\text{min}$), the branching instability that locally reduces the primary spacing (when $\lambda>\lambda_\text{max}$) may also be strongly altered by the presence of fluid flow.
Experimental observations of solidifying a Ti-$\SI{47.5}{\atpercent}$Al-$\SI{2}{\atpercent}$Cr-$\SI{2}{\atpercent}$Nb alloy \cite{viardin2020a} alongside with PF simulations of directional dendritic growth in Ti-$\SI{45}{\atpercent}$-Al \cite{viardin2020a,viardin2020b} indicate that fluid flow in the melt, specifically under hypergravity conditions, strongly affects spacing selection.
Here, we compare DNN prediction with these results and show that the reduction of dendritic spacing is also captured by the side-branching mechanism with the DNN method.

\subsubsection{DNN simulations}

The considered thermophysical, processing, and numerical parameters are listed in Table \ref{tab:ti_al_0}. 
In contrast to the previous section, the simulations are initialized with only one primary dendrite in the center of a domain with $h = 1.75\, R_s$, with $R_s=\SI{2.3}{\micro\meter}$, and size $L_x\times L_y = (1.14\times 0.5)\,\si{\milli\meter^2}$, growing in the $x$-direction. Using a moving frame following the tip position, the most advanced needle tip is fixed at a position of $\SI{0.55}{\milli\meter}$ from the bottom of the domain. The domain is periodic in the $y$-direction (horizontally), and at the top and bottom boundary we apply free-slip conditions with $v_x = 0$. The diffusive field is set to $U=1$ at the top boundary, corresponding to the nominal concentration $c_\infty = \SI{45}{\atpercent}$ of the alloy. At the bottom boundary, no-flux conditions with $\partial U/\partial x = 0$ apply.
Consistently with the corresponding PF study \cite{viardin2020b}, several such simulation were performed using different levels of gravity from $g=-20\,g_0$ to $g=+15\,g_0$, again with positive $g$ corresponding to gravitational forces in the same direction as the primary dendrite growth and temperature gradient.

Another important difference with previous simulations is the presence of side-branching. 
Using a similar approach as in previous implementations \cite{tourret2013a,tourret2016}, new branches perpendicular to the parent dendrite are periodically generated at a distance $l_\text{sb}$ behind the dendrite tip, every time the tip has grown by a distance $l_\text{sb}$. 
The side-branching distance $l_\text{sb}$ of every branch is randomized by adding a random distance $\delta l_\text{sb}$ with range $\left[-\Delta l_\text{sb}/2,+\Delta l_\text{sb}/2\right]$ for each branch independently. Both average side-branching distance and random fluctuation are user-input parameters. 
As long as the distance between side-branches is short enough to induce growth competition among them, this approach was found to be relatively insensitive to selected branching parameters \cite{tourret2013a} and to reproduce scaling laws for experimentally measured dendrite envelopes \cite{tourret2016}.
Here, the side-branching frequency was set at $l_\text{sb}/R_s= \num{7+-1.5}$.

These simulations were performed on a single Nvidia RTX 2080Ti GPU. 
With about 36\,000 grid points for each run, simulations with moderate gravity strength ($|g|<10g_0$) were performed in $\num{2}$ to $\num{3.5}$ days, in contrast to PF simulations lasting approximately a month. 
At higher gravity strength ($|g|>10g_0$), numerical stability required a decrease of the time step, which resulted in those simulations lasting up to two weeks.

\subsubsection{Results and discussion}

\begin{figure*}[t!]
\centering
\includegraphics[width=\textwidth]{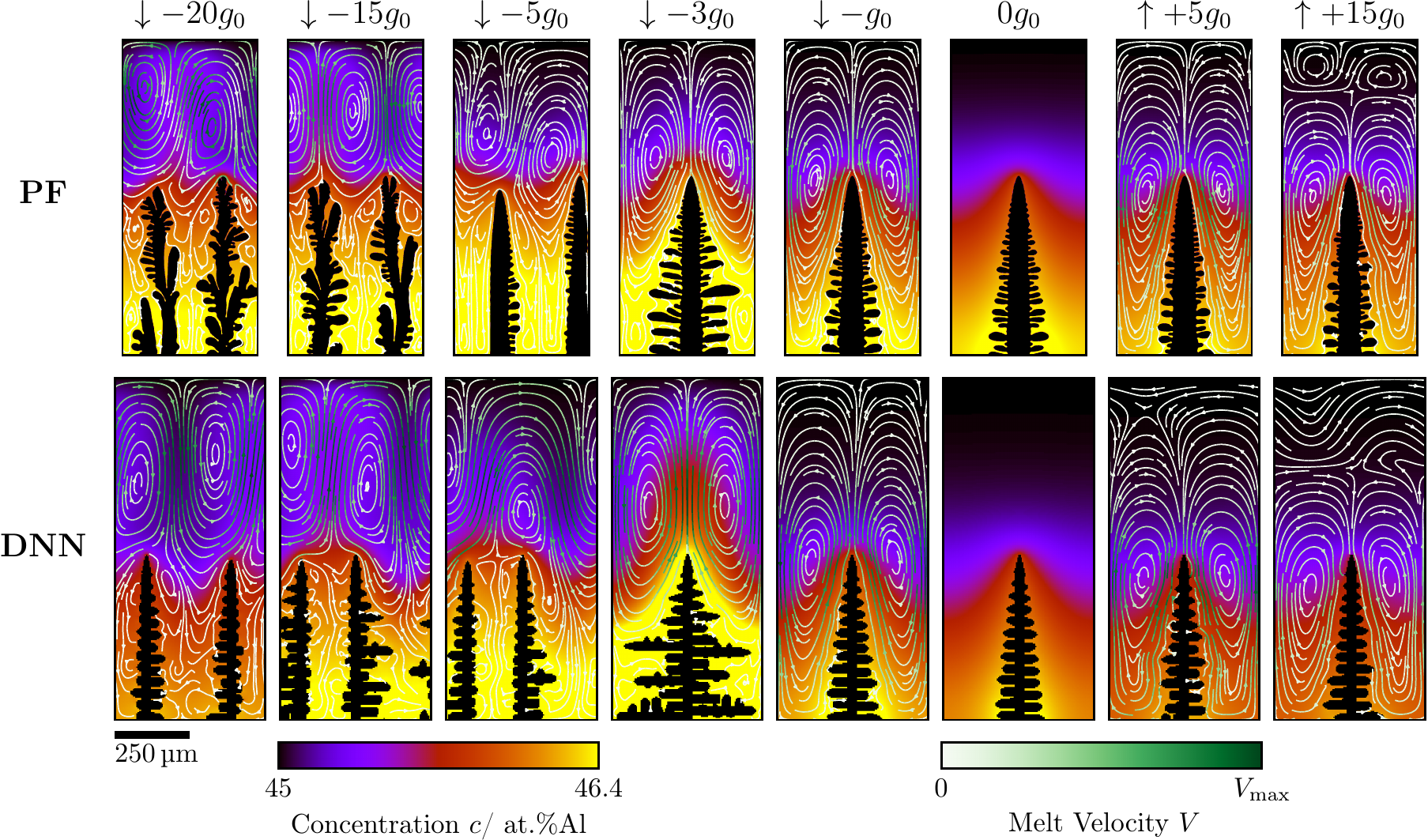}
\caption{Final states of PF \cite{viardin2020b} and DNN simulations at $t=\SI{200}{\second}$ of directional dendritic solidification of Ti-$\SI{45}{\atpercent}$-Al in a horizontally periodic domain. The color map represents the concentration of Al solute, while the melt flow is shown with streamlines colored in shades of green representing the velocity amplitude at each gravity level, respectively. The maximum velocities $V_\text{max}$ from left to right are:  $\{\num{28.0};\,\num{28.9};\,\num{7.1};\,\num{10.6};\,\num{1.0};\,\text{---};\,\num{2.3};\,\num{7.0}\}\times\SI{e-5}{\meter\per\second}$.
\label{fig:ti_al_0}}
\end{figure*}

Fig.\,\ref{fig:ti_al_0} illustrates the final states of the DNN simulations at $t=\SI{200}{\second}$ (bottom) in comparison with the PF results from Ref.~\cite{viardin2020b} (top). 
Tip-splitting events and drifting of the dendrites are not captured by the DNN model, in which the needles have a fixed growth direction and lateral position.
The reduction of primary dendritic spacing for gravity conditions $g<-3g_0$ was nonetheless predicted. 
At $g=-5g_0, -15g_0, -20g_0$ tertiary branches emerge and effectively reduce the spacing.

The overall flow patterns in PF and DNN match qualitatively. At $g=5g_0$ and $g=15g_0$, the DNN simulations develop a lateral flow near the top boundary that differs noticeably with results from PF simulations. This behavior could be rooted in the slightly larger domain size of the DNN simulations with $L_x\times L_y = (0.5\times 1.14)\,\si{\milli\meter^2}$ opposed to $L_x\times L_y = (0.45\times 1.05)\,\si{\milli\meter^2}$. The reason might also be that lateral flow (DNN) or vortices (PF), once established, are not easily disrupted. In any case, the flow velocities near the top boundary are very low compared to the flow close to the dendrite region, such that we can assume this difference to be insignificant.

\section{Oscillatory growth of nickel-based superalloy}
\label{sec:nickel}

In a recent experimental study, solutal buoyant flow was directly observed via in-situ X-ray radiography during directional solidification of a CMSX-4 superalloy \cite{reinhart2020}. 
The effect of the melt flow was evidenced by tracking the dendritic tip growth velocities. 
Depending on the applied cooling rate, tip velocities were observed to exhibit oscillations. 
This oscillatory behavior remains to be simulated and explained in details. 
The current DNN model, which considers a model binary alloy in two dimensions, is not expected to entirely reproduce the complex multicomponent solute interactions in a three-dimensional sample.
Nevertheless, here we show that using a careful pseudo-binary alloy approximation, DNN simulations reproduce this oscillatory growth behavior.

\subsection{Pseudo-binary alloy surrogate}
\label{sec:cmsx4}

We consider the nominal composition of CMSX-4 as listed in Table \ref{tab:cmsx4_composition}.
\begin{table*}[t!]
\centering
\begin{tabular}{lccccccccccc}
\hline
Element & Cr & Co & W & Al & Mo & Re & Ti & Hf & Ta & Ni\\
\hline
Composition $c/\si{\wtpercent}$ & $\num{6.5}$ & $\num{9.6}$ & $\num{6.4}$ & $\num{5.6}$ & $\num{0.6}$ & $\num{3.0}$ & $\num{1.0}$ & $\num{0.1}$ & $\num{6.5}$ & Balance\\
\hline
\end{tabular}
\caption{Considered nominal composition of the CMSX-4 superalloy \cite{reinhart2020}. \label{tab:cmsx4_composition}
}
\end{table*}
In order to reproduce the oscillatory growth behavior, the first task is to design a pseudo-binary surrogate for the CMSX-4 superalloy in the considered growth conditions, namely a temperature gradient of $G=\SI{4400}{\kelvin\per\meter}$ and a velocity range between $\SI{7.58}{\micro\meter\per\second}$ and $\SI{31.4}{\micro\meter\per\second}$ in experiments, or up to $\SI{60.6}{\micro\meter\per\second}$ in the DNN simulations.
In particular, we aim for a reasonable description of (i) the crystal growth kinetics and (ii) the buoyant flow patterns and velocities.
Regarding the growth kinetics, we consider that the description is acceptable if, for the considered velocity range, the dendrite tip radius of the surrogate alloy matches that of the full CMSX-4 alloy, and the velocity for the onset of constitutional undercooling is also reasonably approximated.
In terms of fluid flow, we want to consider the alloying elements that play the most prominent role in the buoyant flow and approximate their average solutal expansion coefficient.

Starting with the buoyancy consideration, we estimate that the species responsible for the formation of buoyant plumes are the lightest alloying elements, namely aluminum and titanium.
Indeed, chromium, cobalt, and molybdenum are close enough from nickel in weight for their influence on buoyancy to be minor.
Heavier elements, on the other hand, like tantalum, tungsten, and rhenium, may lead to non-negligible buoyant forces, but with a stabilizing effect as the heavier liquid would sink between the primary dendrites \cite{steinbach2009}.

Using the CalPhaD method (software: ThermoCalc, database: TCNI8), we calculated the thermodynamic equilibrium of the full CMSX-4 alloy (excluding minor alloying element hafnium absent from the database) at its liquidus temperature, $T_L\approx1660\,$K.
At this temperature, we verified that aluminum and titanium indeed have the highest solutal expansion coefficients (Eq.~\eqref{eq:solutal_expansion_coefficient}) with $\beta_{\rm Al}\approx 1.35\times10^{-2}$/wt\% and  $\beta_{\rm Ti}\approx 0.75\times10^{-2}$/wt\%, compared to $\beta_{\rm Cr}\approx 0.21\times10^{-2}$/wt\%, $\beta_{\rm Co}\approx 0.09\times10^{-2}$/wt\%, and $\beta_c<0$ for heavier elements Mo, Re, Ta, and W.
These values are consistent and close with those mentioned in the literature \cite{iida1993,schneider1997}

From these considerations, we opted for a surrogate \{A+B\} alloy, where the solute B=\{Al+Ti\} combines Al and Ti contributions, while solvent A represents the other elements.
Its solutal expansion coefficient is approximated as $\beta_c\approx\num{e-2}$/wt\%.
The alloy nominal concentration is taken as $c_\infty=c_\infty^\text{Al}+c_\infty^\text{Ti}=\SI{6.6}{\wtpercent}$.

For the diffusion coefficient $D$, partition coefficient $k$, and liquidus slope $m$, we want to use realistic orders of magnitude, relevant to Al and Ti species, such that the operating state of a steady-state growing dendrite, namely its tip radius $R$ and velocity $V$, matches that expected for the full CMSX-4 alloy at the considered $G$.
In particular, we aim for a good agreement between pseudo-binary surrogate and full CMSX-4 alloy in terms of predictions of the classical Kurz-Giovanola-Trivedi (KGT) model \cite{kurz1986} extended to multicomponent alloys.
We also aim at a good match in terms of onset velocity for constitutional undercooling, such that the considered growth velocity appropriately falls within the dendritic regime.

Diffusivities of aluminum and titanium species in liquid nickel were assessed as $D_\text{Al}=\SI{1.86e-7}{\meter^2\per\second}\times\exp\{\SI{-0.66}{\eV}/(k_BT)\}$ (for a Ni$_{87.5}$Al$_{12.5}$ alloy) \cite{levchenko2017composition} and $D_\text{Ti}=\SI{1.70e-7}{\meter^2\per\second}\times\exp\{\SI{-57.43}{\kilo\joule\per\mol}/(R_gT)\}$ \cite{walbruhl2018atomic}, yielding $D_\text{Al}\approx\SI{2.05e-9}{\meter^2\per\second}$ and $D_\text{Ti}\approx\SI{2.92e-9}{\meter^2\per\second}$ at $T=\SI{1700}{\kelvin}$.
Aluminum being the major alloying element, we chose $D=\SI{2.0e-9}{\meter^2\per\second}$ as a good approximation.

Solute partition coefficients for Al and Ti calculated with CalPhaD for the CMSX-4 alloy at its liquidus temperature are respectively $k_\text{Al}=0.9$ and $k_\text{Ti}=0.46$. In binary Ni-5.6wt\%Al and Ni-1.0wt\%Ti alloys, partition coefficients at their respective liquidus temperatures are $k_\text{Al}=0.90$ and $k_\text{Ti}=0.64$.
For the binary surrogate approximation, we used an intermediate value, closer to that of aluminum, with $k=0.8$.

\begin{table*}[t]
\centering
\begin{tabular}{lcrl}
\hline
Property & Symbol & Value & Unit \\
\hline
 Nominal composition & $c_\infty$ & $6.6$ & $\si{\wtpercent}$ \\
 Liquid Diffusivity & $D_l$ & $\num{2e-9}$ & $\si{\meter^2\per\second}$ \\
 Partition coefficient & $k$ & $0.8$ & \\
 Liquidus slope & $m$ & $-25$ & $\si{\kelvin\per\wtpercent}$ \\
 Solutal expansion coefficient & $\beta_c$ & $\num{e-2}$ & $\si{\per\wtpercent}$ \\
 Gibbs-Thomson coefficient & $\Gamma$ & $\num{2.49e-7}$ & $\si{\kelvin\meter}$ \\
 Kinematic viscosity & $\nu$ & $\num{5.8e-7}$ & $\si{\meter^2\per\second}$ \\
 Interfacial energy anisotropy & $\epsilon$ & $\num{1.2e-2}$ & \\
 Tip selection parameter & $\sigma$ & $\num{0.08}$ & \\
\hline
Temperature gradient & $G$ & $\num{4.4e3}$ & $\si{\kelvin\per\meter}$ \\
Cooling rate & $\dot T$ & $\{-2; -4; -8.3; -11; -13; -16\}$ & $\si{\kelvin\per\minute}$ \\
\hline
FIF integration radius & $r_i$ & $\numrange{5.4}{8.5}$ &$R_s$ \\
Parabola truncation radius & $r_\text{max}$ & $\numrange{5.4}{8.5}$ &$R_s$ \\
Upwind parameter & $\omega_\text{up}$ & $0.9$ & \\
Successive Over Relaxation parameter  & $\omega_\text{SOR}$ & $1.1$ & \\
SOR residual required for convergence & $\overline{r}_\text{SOR}$ & $\num{e-3}$ & \\
Time step safety factor & $K_{\Delta t}$ & $\num{0.15}$ & \\
\hline
\end{tabular}
\caption{Material and processing parameters used int he DNN simulation of the directional solidification of the CMSX-4 surrogate alloy (see Sec.\,\ref{sec:cmsx4} for sources) and numerical parameters (see ref. \cite{tourret2019} for computational details) \label{tab:cmsx4_0}}
\end{table*}

\begin{table*}[t]
\centering
\begin{tabular}{lcccc}
\hline
Cooling Rate $\dot T$ & Grid Spacing $h$ & Domain Size $N_x\times N_y$ & Domain Size $L_x\times L_y$ & Initial PDAS\\
\hline
$\SI{-2}{\kelvin\per\minute}$ & $\SI{1.5}{\dnnr}=\SI{10.4}{\micro\meter}$ & $766\times414 $ &        $\SI{8}{\milli\meter} \times\SI{4.3}{\milli\meter} $ & $\SI{239}{\micro\meter}$ \\ 
$\SI{-4}{\kelvin\per\minute}$ & $\SI{2.125}{\dnnr}=\SI{10.4}{\micro\meter}$ & $766\times414 $ &    $\SI{8}{\milli\meter} \times\SI{4.3}{\milli\meter} $ & $\SI{239}{\micro\meter}$ \\ 
$\SI{-8.3}{\kelvin\per\minute}$ & $\SI{1.85}{\dnnr}=\SI{6.3}{\micro\meter}$ & $766\times510 $ &     $\SI{4.8}{\milli\meter} \times\SI{3.2}{\milli\meter} $ & $\SI{213}{\micro\meter}$ \\
$\SI{-11}{\kelvin\per\minute}$ & $\SI{1.6}{\dnnr}=\SI{4.7}{\micro\meter}$ & $1022\times638 $ &      $ \SI{4.8}{\milli\meter} \times\SI{3}{\milli\meter} $ & $\SIrange[range-units=brackets]{272}{375}{\micro\meter}$ \\
$\SI{-13}{\kelvin\per\minute}$ & $\SI{1.5}{\dnnr}=\SI{4}{\micro\meter}$ & $1022\times638 $ &         $\SI{4.1}{\milli\meter} \times\SI{2.6}{\milli\meter} $ & $\SI{235}{\micro\meter}$ \\
$\SI{-16}{\kelvin\per\minute}$ & $\SI{1.35}{\dnnr}=\SI{3.3}{\micro\meter}$ & $1022\times1022 $  & $\SI{3.4}{\milli\meter} \times\SI{3.4}{\milli\meter} $ & $\SI{239}{\micro\meter}$ \\  
\hline
\end{tabular}
\caption{Grid spacings and domain sizes of the DNN simulations at different cooling rates $\dot T$.\label{tab:cmsx4_numerical}}
\end{table*}

\begin{figure}[t!]
\centering
\includegraphics[width=\columnwidth]{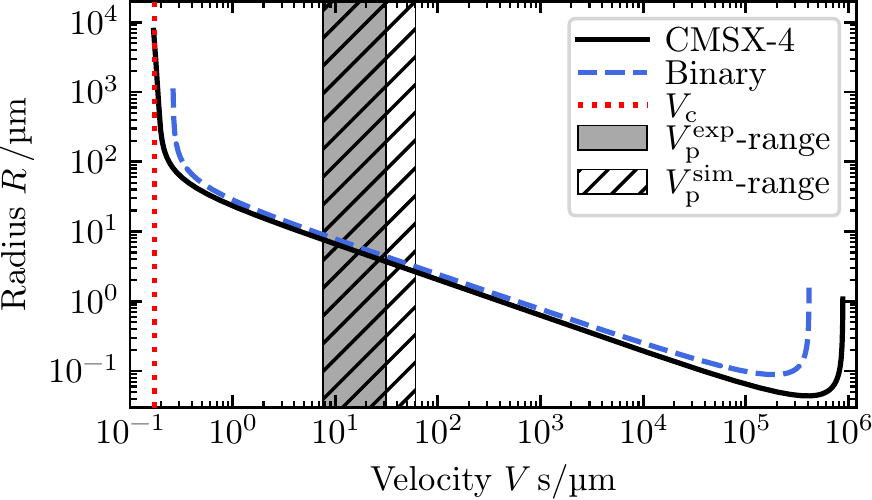}
\caption{KGT model prediction of dendrite tip radius versus velocity for multicomponent CMSX-4 (Supplementary Material, Table~\ref{tab:parameters}) and binary surrogate alloy (Table~\ref{tab:cmsx4_0}).
\label{fig:nickel_based_kgt_0}}
\end{figure}

CalPhaD-calculated liquidus slopes respective to Al and Ti in the CMSX-4 alloy are $m_\text{Al}\approx\SI{-13.1}{\kelvin\per\wtpercent}$ and $m_\text{Ti}\approx\SI{-20.3}{\kelvin\per\wtpercent}$.
In binary Ni-5.6wt\%Al and Ni-1.0wt\%Ti alloys, liquidus slopes are $m_\text{Al}\approx\SI{-4.92}{\kelvin\per\wtpercent}$ and $m_\text{Ti}\approx\SI{-9.77}{\kelvin\per\wtpercent}$.
However, we found that using such values leads to a notable discrepancy between pseudo-binary and full CMSX-4 alloy in terms of KGT-predicted tip radius $R(V)$ and constitutional undercooling velocity $V_c$.
Hence, we treated $m$ as an adjustable parameter to better match $R(V)$ and $V_c$.
We used a simple extension of the KGT model \cite{kurz1986} to multicomponent alloys \cite{rappaz1989,rappaz1990} by adding up solutal contributions of the different alloying elements (see Supplementary Material).
This formulation neglects cross-species interactions \cite{hunziker2001theory}, which is typically acceptable for relatively dilute solute species, and yields predictions of planar interface stability limits consistent with this assumption \cite{coates1968solid}.
As depicted in Fig.\,\ref{fig:nickel_based_kgt_0}, resulting KGT calculations lead to a CMSX-4 onset of constitutional undercooling at a velocity $V_c\approx\SI{1.75e-7}{\meter\per\second}$ (see details and parameters in Section~\ref{sm_kgt} of the Supplementary Material).
%
In order to match this velocity for the parameters considered here, using the classical binary criterion $V_\text{c}=DGk/\left[m(1-k)c_\infty\right]$ \cite{tiller1953redistribution, mullins1964stability} for the temperature gradient $G=\SI{4400}{\kelvin\per\meter}$ of the experiments \cite{reinhart2020}, we obtain a liquidus slope $m\approx\SI{-25}{\kelvin\per\wtpercent}$.
Although this value is higher than CalPhaD-calculated values, it remains within the same order of magnitude, and we decided to use it for the surrogate alloy, since it leads to a good approximation of the full CMSX-4 alloy in the KGT-predicted $R(V)$ curve in the considered velocity range (Fig.\,\ref{fig:nickel_based_kgt_0}).

Remaining parameters, namely kinematic viscosity, Gibbs-Thomson coefficient, and interface energy anisotropy were estimated for pure Ni.
We considered a kinematic viscosity $\nu=\SI{5.8e-7}{\meter^2\per\second}$ using dynamic viscosity and density values determined experimentally for pure Ni in Refs \cite{sato2005} and \cite{cagran2007}, respectively.
For the Gibbs-Thomson coefficient, we used $\Gamma=\gamma_0T_M/L_f=\SI{2.49e-7}{\kelvin\meter}$, considering pure Ni melting temperature $T_M=\SI{1728}{\kelvin}$ and latent heat of fusion $L_f=\SI{2.08e9}{\joule\per\meter^3}$ calculated with CalPhaD (TCNI8), and an interface excess free energy $\gamma_0\approx\SI{0.3}{\joule\per\meter^2}$, consistent with several independent calculations using molecular dynamics (capillary fluctuation method) between $\num{0.271}$ and $\SI{0.364}{\joule\per\meter^2}$\cite{hoyt2003atomistic, jiang2008size, asadi2015two}.
The fourfold interface free energy anisotropy was considered $\epsilon=0.012$, which corresponds, for a one-sided model in 2D \cite{barbieri1989}, to a tip selection parameter $\sigma\approx0.08$.

Assumptions made here in the construction of a pseudo-binary CSMX-44 surrogate are arguably approximate, specific to the problem that we aim to simulate, and not intended as a general procedure for pseudo-binary approximations of complex multicomponent alloys.
Nonetheless, we will see in the following subsections that this simple description is sufficient to reproduce and hence investigate the experimentally-observed oscillatory growth regime.

\subsection{DNN simulations}

Table \ref{tab:cmsx4_0} summarizes the material, processing and numerical parameters used in the DNN simulations. 
We simulated the directional solidification of the surrogate alloy for six different cooling rates from $-2$ to $\SI{-16}{\kelvin\per\minute}$.
The three lowest cooling rates of $-2$, $-4$, and $\SI{-8.3}{\kelvin\per\minute}$ correspond to the experimental conditions.
The grid spacing $h$ was set between 1.35 and 2.125 times the steady tip radius $R_s$, while ensuring that $h \leq D/(10V_s)$ in order to provide an appropriate spatial description of solute gradients.
Table \ref{tab:cmsx4_numerical} summarizes the corresponding numerical parameters. 
The domain was initialized with an array of between 6 and 22 evenly-spaced primary dendrites, with their tips initially located at the liquidus temperature.
The growth of the dendritic arrays was simulated for a physical time of between $\SI{2.5}{\minute}$ (for $\dot T=\SI{-16}{\kelvin\per\minute}$) and $\SI{28}{\minute}$ (for $\dot T=\SI{-2}{\kelvin\per\minute}$).
In order to assess the effect of primary spacing on the oscillatory growth behavior, we also performed simulations at $\dot T=\SI{-11}{\kelvin\per\minute}$, using different initial spacings of 272, 300, and $\SI{375}{\micro\meter}$.
In all simulations, boundary conditions were similar as those used in Sec. \ref{subsec:verification_al-cu} and side-branching was not enabled.

All simulations were carried out on a single Nvidia RTX3090 GPU. 
Simulation times ranged between 6.5~days (for $\SI{28}{\minute}$ at $\SI{-8.3}{\kelvin\per\minute}$) and 24~days (for $\SI{12}{\minute}$ at $\SI{-11}{\kelvin\per\minute}$).
As a representative example, the simulation discussed later in Fig.\,\ref{fig:nickel_based_1}, corresponding to $\SI{5}{\minute}$ of cooling at $\SI{-13}{\kelvin\per\minute}$, was performed in 14~days.

\subsection{Results and discussion}

\begin{figure}[b!]
\centering
\includegraphics[width=\columnwidth]{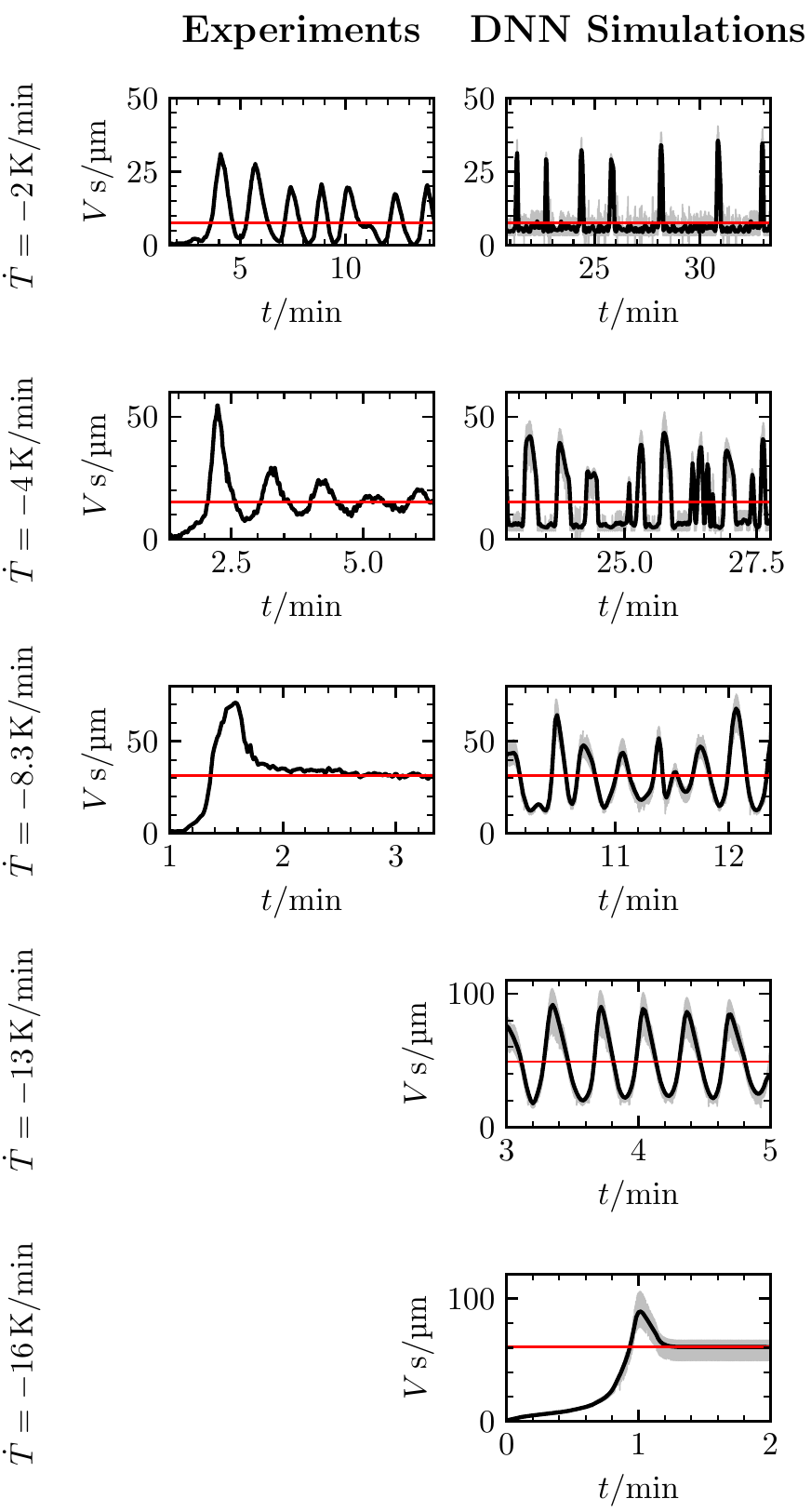}
\caption{Experimentally measured \cite{reinhart2020} solidification velocities (left) and corresponding DNN predicted velocities (right) at different cooling rates. 
Gray and black lines respectively correspond to the raw and time-averaged data (using a moving average over $\SI{1}{\second}$).
Horizontal red lines represent the steady-state tip velocity, i.e. the equivalent pulling velocity $V_p=|\dot T|/G$.
\label{fig:nickel_based_0}}
\end{figure}

Fig.\,\ref{fig:nickel_based_0} shows a side-by-side comparison of the experimentally measured \cite{reinhart2020} solidification velocities and the velocities predicted by the DNN model. 
For each cooling rate, $V(t)$ from experiments and simulations are represented using the same time and velocity scales, with the equivalent pulling velocity ($V_p=|\dot T|/G$) marked with a red horizontal line.
Experimental velocities correspond to the tip of one central dendrite (see Fig.~7 and corresponding discussion in Ref.~\cite{reinhart2020}).
Simulation results correspond to the velocity of a single arbitrarily-chosen dendrite tip that did not get eliminated throughout the simulation. 
Similar plots showing $V(t)$ for every single dendrite in each simulation are provided in the Supplementary Material (Fig.~\ref{fig:nickel_based_velocities_R}), showing that the behaviors illustrated in Fig.\,\ref{fig:nickel_based_0} are representative of those across the entire dendritic array.

As discussed in Ref.~\cite{reinhart2020}, during directional solidification experiments, velocity oscillations were identified that were sustained over tens of minutes, with an oscillation period of about 80~seconds when cooling at $\SI{-2}{\kelvin\per\minute}$.
In other experiments with faster cooling rates, oscillations of comparable period were progressively damped as the cooling rate was increased.

In the results from DNN simulations, low cooling rates ($|\dot T|\leq\SI{4}{\kelvin\per\minute}$) lead to growth fluctuations, but they appear quite random, with single sharp spikes.
The growth regime progressively transitions to a more periodic behavior at faster cooling rates.
The sustained oscillatory growth and its attenuation when increasing the cooling rate is also observed in DNN simulations (Fig.\,\ref{fig:nickel_based_0}), however for a higher cooling rate than in the experiments. 
Sustained oscillations, experimentally identified at $\dot T=\SI{-2}{\kelvin\per\minute}$, appear in the simulations around $\SI{-13}{\kelvin\per\minute}$, with a period of about 20~second.

\begin{figure}[b!]
\centering
\includegraphics[width=\columnwidth]{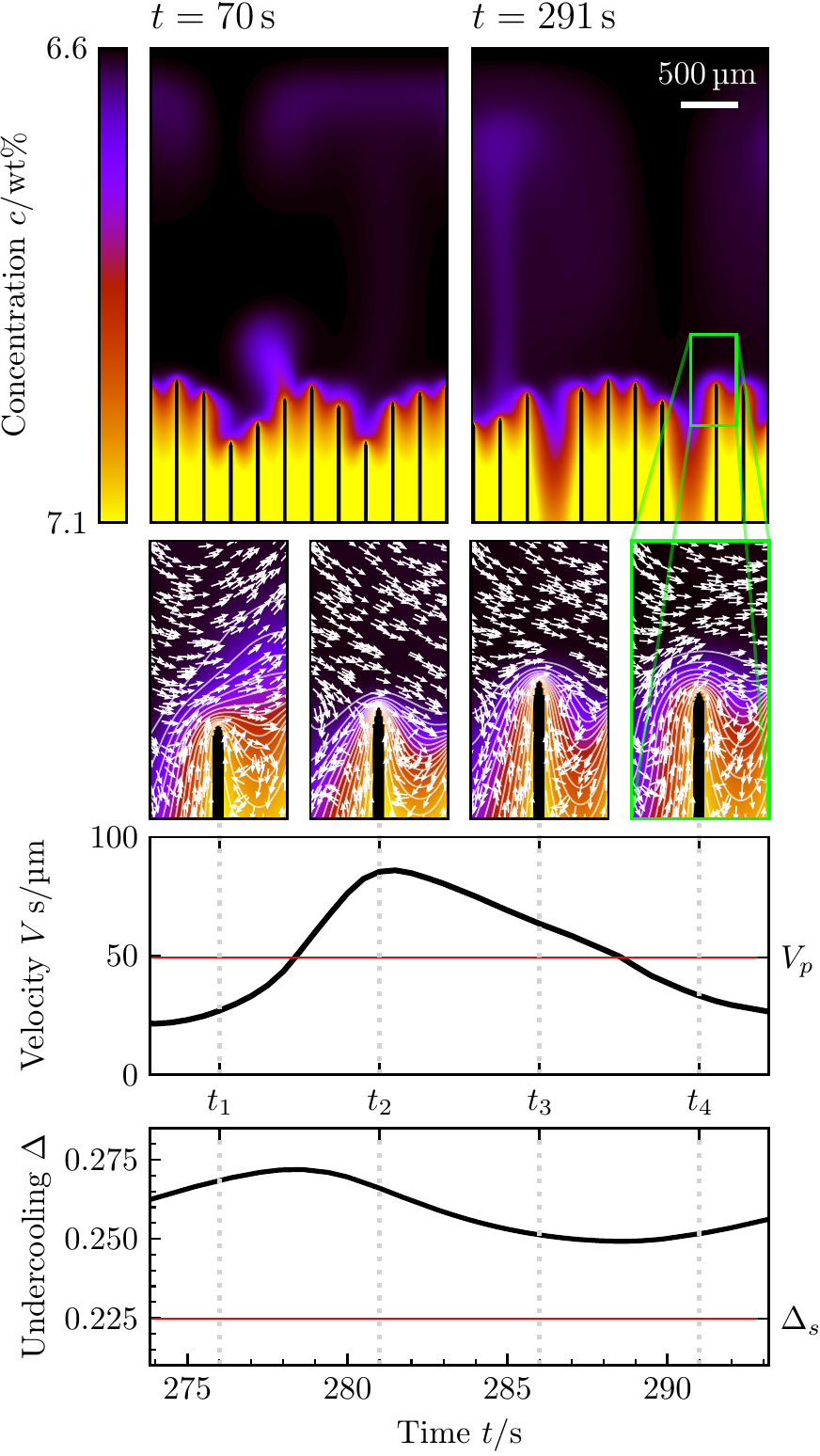}
\caption{Concentration fields of the full simulation domain (top) at $t=\SI{70}{\second}$ and $t=\SI{291}{\second}$ for the cooling rate $\dot T=\SI{-13}{\kelvin\per\minute}$, with enhanced snapshots of the area marked by the green rectangle and its corresponding tip velocity and undercooling during a single oscillation period (below).
The four panels are aligned with the corresponding times $t_1$, $t_2$, $t_3$, and $t_4$ in the bottom plots of the tip velocity and undercooling. 
Arrows indicate the flow direction and the color map and iso-contours represent the solute concentration $U$.
\label{fig:nickel_based_1}
}
\end{figure}

Fig.\,\ref{fig:nickel_based_1} illustrates the behavior of the flow pattern when oscillations occur for $\dot T=\SI{-13}{\kelvin\per\minute}$. 
The dendrite marked by the green rectangle in the full domain on the right side, is shown at four time steps during one oscillation period. 
Although the alternating flow patterns are complex when approached at the scale of the entire domain, clear trends emerge when looking at the overall flow direction (white arrows) surrounding a given dendrite tip. 
The dendrite grows at its lowest velocity ($t_1$, $t_4$) when the fluid exhibits a strong upward current, thus locally depleting the region surrounding the tip in solute. 
The tip velocity is maximal ($t_2$) when the flow has a strong downward component, feeding the tip in solute. 
At intermediate velocities ($t_3$) the liquid predominantly flows laterally, which is known to lead to a tip growth velocity comparable to that in the absence of convection \cite{tong2001phase, jeong2001, badillo2007growth,sakane2018three}.
Overall, the dendrite tip velocity oscillates around the equivalent pulling velocity $V_p$.
Meanwhile, dendrites within the array are still interacting with each other via the solute field, such that the tip undercooling oscillates above the theoretical undercooling $\Delta_s$ for a free (i.e. isolated) dendrite.

In order to assess the range of cooling rates at which oscillations occur, we estimated the average flow velocity $\overline V$ in the final stage of each simulation.
To do so, we extracted the spatial average of the amplitude of the velocity field in the liquid for five different time steps, within one oscillation period (or over $\SI{16}{\second}$ in the late stages of the simulations when oscillations are absent), and used the average of those five values as an approximate velocity over space and time.
For completeness, the five snapshots used for each simulations are illustrated in the Supplementary Material (Fig.~\ref{fig:velocit_average}).
Results of this analysis, summarized in Table~\ref{tab:cmsx4_averageV}, clearly identify that oscillations occur when the average flow velocity $\overline V$ is close to the equivalent pulling velocity $V_p$.
Indeed, when $\overline V/V_p<1$ oscillations are damped, when $1<\overline V/V_p<2$ oscillations are sustained, and for higher $\overline V/V_p$ the growth behavior becomes increasingly more erratic.
While the current estimation of $\overline V$ is arguably approximate, this analysis unambiguously demonstrates that oscillations occur when the flow velocity and the growth velocity are of the same order of magnitude.

\begin{table}[b]
\centering
\begin{tabular}{rccccc}
\hline
& $\dot T$ & $V_p$ & $\overline V$ &  $\overline V/V_p$ & Oscillations \\
& $\si{\kelvin\per\minute}$ & $\si{\micro\meter\per\second}$ & $\si{\micro\meter\per\second}$ &  & \\
\hline
&     $-2$   & 7.6 & 167.2 & 22.0 & Spikes \\
&     $-4$   & 15.2 & 202.2 & 13.3 & Spikes \\
&    $-8.3$ & 31.4 & 119.3 & 3.79 & Intermediate  \\
(a) &$-11$  & 41.7 & 25.4 & 0.61 & Damped \\
(b) &$-11$  & 41.7 & 14.6 & 0.35 & Damped \\
(c) &$-11$  & 41.7 & 75.5 & 1.81 & Sustained \\
&    $-13$  & 49.2 & 76.8 & 1.56 & Sustained \\
&    $-16$  & 60.6 & 0.61 & 0.01 & Damped \\
\end{tabular}
\caption{Average flow velocities $\overline V$ (see text and Fig.~\ref{fig:velocit_average} of the Supplementary Material) and equivalent pulling velocity $V_p$ for the different DNN simulations. The three cases labeled (a), (b), and (c) at $\dot T=\SI{-11}{\kelvin\per\minute}$ correspond to the simulations illustrated in Fig.~\ref{fig:nickel_based_2}.
\label{tab:cmsx4_averageV}
}
\end{table}

The discrepancy in cooling rate leading to oscillations between experiments and simulations may be attributed to assumptions made in the pseudo-binary approximation of the CMSX-4 alloy (see Sec.\,\ref{sec:cmsx4}), as well as dimensionality, as pointed out already in Sec.\,\ref{sec:alcu_resu}.
Indeed, since the flow velocity is overestimated in the 2D simulations \cite{jeong2001,isensee2020}, the range of cooling rates with $\overline V\approx V_p$ occurs at higher $V_p$, i.e. at higher $|\dot T|$.
The difference in oscillation period likely stems from the fact that the oscillations appear at a higher cooling rate, and therefore is also due to dimensionality and surrogate alloy approximations.
Preliminary observations suggest an increase of oscillation frequency with cooling rate. 
However, since the range of cooling rate resulting in sustained oscillations is relatively narrow, this difference is limited ($\approx\SI{15}{\percent}$ increase from $\SI{-11}{\kelvin\per\minute}$ to $\SI{-13}{\kelvin\per\minute}$), and one may expect a greater influence of alloy parameters (in particular $\nu$, $D$, and $\beta_c$).
Further ongoing parametric studies on a broader range of alloys will clarify the influence of these parameters in the oscillation frequency. 

\begin{figure*}[t!]
\centering
\includegraphics[width=\textwidth]{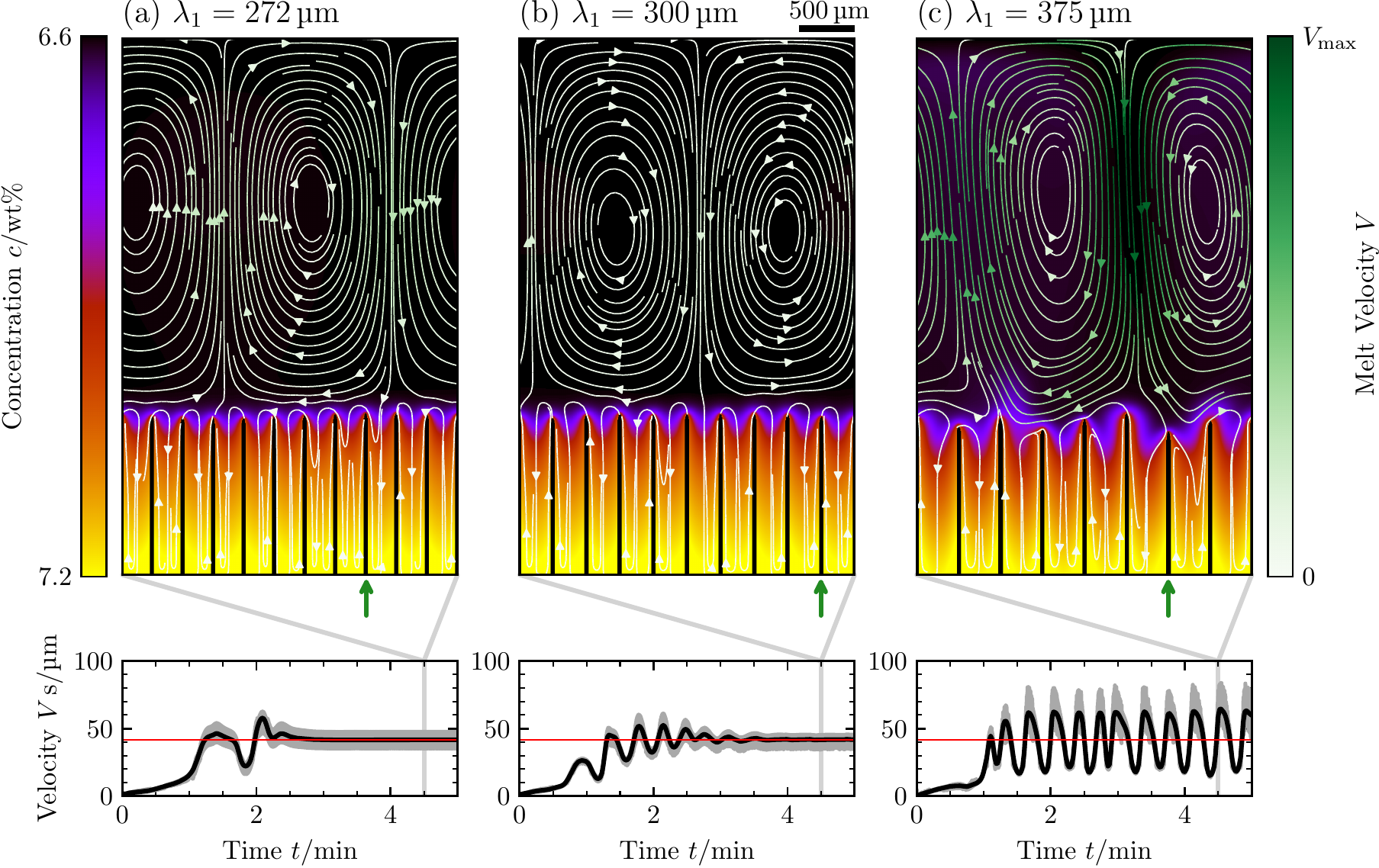}
\caption{DNN-predicted flow patterns (top) and tip velocities $V$ (bottom) at $\dot T=\SI{-11}{\kelvin\per\minute}$ at $t=\SI{4.5}{\minute}$ with different primary dendrite arm spacing $\lambda_1$. The streamlines represent the fluid flow with maximum velocity $V_\text{max} = \num{102}~V_p = \SI{4.3e-3}{\meter\per\second}$. The illustrated tip velocities correspond to the dendrites indicated by green arrows. The gray and black curves correspond to raw and smoothed data, respectively. The red horizontal lines represent the theoretical steady state tip growth velocity, i.e. the equivalent pulling velocity $V_p$. Plots of the velocities of all needles can be found in Fig.~\ref{fig:nickel_based_velocities_N} of the Supplementary Material.\label{fig:nickel_based_2}}
\end{figure*}

In addition to its dependence upon cooling rate, we also found that the oscillatory behavior was strongly dependent upon primary dendritic spacing. 
In Fig.\,\ref{fig:nickel_based_1}, for instance, the oscillatory growth occurs after two initially set primary dendrites (present in the top-left snapshot at $t=\SI{70}{\second}$) were eliminated.
In order to assess the influence of spacing, we performed simulations at a cooling rate $\dot T=\SI{-11}{\kelvin\per\minute}$ using different primary spacings within a range where no elimination event occurs, i.e. namely for $\lambda_1=272$, 300, and $\SI{375}{\micro\meter}$.
Fig.\,\ref{fig:nickel_based_2} illustrates the resulting concentration fields at $t=\SI{4.5}{\minute}$, as well as the velocity evolution of one needle (marked with a green arrow) in each simulation.
The velocities of all needles (provided in Fig.~\ref{fig:nickel_based_velocities_N} of the Supplementary Material) exhibit a similar behavior as the ones highlighted here. For (a) $\lambda=\SI{272}{\micro\meter}$, dendrites are close to one another.
Initial transient oscillations are quickly damped and the dendrites finally grow together at the steady-state velocity $V_p=\SI{41.7}{\micro\meter\per\second}$.
As the spacing gets larger (b), damping of the oscillations occurs over a longer time during which several oscillation periods are noticeable.
Ultimately (c), higher spacings allow stronger convective currents, leading to a sustained oscillatory growth of the dendritic array. 
Values of the average fluid velocity estimated in these three simulations (see Table~\ref{tab:cmsx4_averageV}), are consistent with our observation that oscillations occur when $\overline V$ is higher yet close to $V_p$, with $\overline V<V_p$ in both damped cases (a) and (b), and $\overline V/V_p\approx1.86$ in the sustained case (c).

Our interpretation of the effect of spacing on the occurrence of oscillations is the following. 
First, for the oscillations to take place, sufficient fluid flow must be allowed between the primary dendrites.
Second, since the oscillations of adjacent primary dendrites are out-of-phase with each other, a lateral symmetry-breaking must arise, leading to a two-dimensional composition profile.
When the primary spacing is low, fluid flow is strongly limited between the dendrites. 
This is illustrated in Fig.~\ref{fig:velocities_line_sample} of the Supplementary Material, which represents the vertical ($x$) component of the velocity averaged over the entire liquid region at a given height ($\overline{|V_x|}$) at five different times ($t=246$, 259, 273, 286, and $\SI{300}{\second}$), and clearly shows that the resulting velocity is much higher for the highest spacing case of Fig.~\ref{fig:nickel_based_2}c.
When the primary spacing is low, the composition field between the dendrites and ahead of the solidification front also remains relatively close to a one-dimensional profile ahead of a planar front. 
At higher spacings, lateral composition gradients ($\partial c/\partial y$) can develop, which lead to the symmetry breaking, to the development of a two-dimensional composition field, and to the emergence of oscillations.
This is illustrated in Figs~\ref{fig:concentrations_line_sample} and \ref{fig:concentrations_line_sample_horizontal} of the Supplementary Material, respectively showing the composition profile along vertical lines located at the center between two primary dendrites at different times (Fig.~\ref{fig:concentrations_line_sample}), and the composition profile along horizontal lines in the liquid region at $x-x_{tip}=\SI{141}{\micro\meter}$ and $\SI{423}{\micro\meter}$ at five different times (Fig.~\ref{fig:concentrations_line_sample_horizontal}).
These plots show that for $\lambda_1=\SI{272}{\micro\meter}$ (Fig.~\ref{fig:nickel_based_2}a) and $\lambda_1=\SI{300}{\micro\meter}$ (Fig.~\ref{fig:nickel_based_2}b) the composition profile is essentially one-dimensional, with $(c-c_\infty)/c_\infty$ remaining below $\SI{0.1}{\percent}$ in the liquid ahead of the dendrite tips, while the composition profile for $\lambda_1=\SI{375}{\micro\meter}$ (Fig.~\ref{fig:nickel_based_2}c) exhibits significantly higher composition gradients in the $y$ direction.
The transition from damped to sustained oscillations seems to occur when the primary spacing $\lambda_1$ is between 6 and 8 times $D/V_p$, since at this cooling rate the steady-state diffusion length is $D/V_p\approx\SI{48}{\micro\meter}$.
Additional simulations should clarify how this threshold changes within a broader range of alloys and processing parameters --- and whether the diffusion length is the appropriate length scale with which to compare in presence of appreciable convection. 

It is worth mentioning that Fig.\,\ref{fig:nickel_based_2} illustrates special cases in which the dendritic array is perfectly regular and no elimination occurs. 
In general cases, as those depicted in Figures~\ref{fig:nickel_based_0} and \ref{fig:nickel_based_1}, a symmetry breaking due to the occurrence of elimination events leads to more complex overall array dynamics.
Hence, for low spacings, expected to lead to complete damping of the oscillations, elimination events may lead to an increase of average primary spacing, and consequently to an oscillatory growth behavior. 
This is illustrated, for instance, in additional simulations, using different initial primary spacings at $\dot T=\SI{-13}{\kelvin\per\minute}$ presented in the Supplementary Material (Fig.~\ref{fig:nickel_based_velocities_N}).

In summary, the current results confirmed that a buoyancy-induced oscillatory growth behavior, observed in experiments \cite{reinhart2020}, may occur across a narrow range of cooling rates when the average flow velocity is close to the average growth velocity, and suggested that primary dendritic spacings play a prominent role in the resulting oscillations being sustained or damped. 
Ongoing investigations should provide a deeper understanding of the mechanism, e.g. establishing relevant scaling laws for the resulting oscillation period and amplitude.
Further applications of the model to polycrystalline growth with nucleation \cite{geslin2021dendritic, chen2021dendritic} should also allow the simulation of segregated channels and freckle formation. However, an extension of the current model would remain required to treat potential remelting and fragmentation events in the segregated channels, as well as the buoyant motion of stray crystals.

\section{Summary and conclusions}
\label{sec:conclusion}

We presented a two-dimensional implementation of the dendritic needle network (DNN) model for directional solidification of binary alloys with buoyant melt flow. 
Results of the model regarding the selection of primary dendritic spacings in Al-$\SI{4}{\atpercent}$-Cu and Ti-$\SI{45}{\atpercent}$-Al alloys under various gravity conditions are consistent with previously reported phase-field and experimental data.
Scaling laws for the lower spacing limit $\lambda_\text{min}$ for upward and downward flows were reproduced \cite{steinbach2009}.
Spacing reduction via side-branching in DNN simulations reasonably mimic tip-splitting events expected in presence of strong gravity in direction opposite to the growth \cite{viardin2020b}.

We simulated the experimentally observed oscillatory growth behavior in nickel-based single-crystal CMSX-4 alloy \cite{reinhart2020}.
To do so, we considered a surrogate binary alloy, derived from simple assumptions using CalPhaD calculations and classical solidification theories, namely matching predictions of constitutional undercooling criterion and KGT model. 
Oscillatory growth velocities were reproduced, however at cooling rates slightly higher than identified in experiments. 
The discrepancy is mainly attributed to dimensionality, since flow velocities are usually overestimated in two-dimensional simulations \cite{jeong2001,isensee2020}. 
Our results confirmed that the oscillatory growth behavior is closely linked to the buoyant flow in the liquid phase, that it occurs over a narrow range of cooling rates (i.e. growth velocity) for a given temperature gradient, and that the oscillatory behavior strongly depends on the primary dendritic spacing.

In summary, we used a new model to (i) gain new fundamental insights into an important yet still incompletely understood aspect linking materials processing and microstructure, namely during solidification in the presence of fluid flow, and (ii) validate those insights by a direct comparison of modeling predictions and state-of-the-art in situ imaging experiments in a technologically important application – namely directional solidification of a single-crystal Ni-based superalloy. 

Ongoing and future investigations following on this study include applications of the model to a broader range of experiments (e.g. Ref.~\cite{gibbs2016situ}) as well as three-dimensional applications \cite{isensee2020}.
Among other things, the computationally efficient DNN simulations should allow further study of the dependence of dendrite growth kinetics upon the surrounding flow strength and direction \cite{badillo2007growth,sakane2018three}.
A deeper investigation into oscillatory growth behaviors during directional solidification, scanning a wider range of alloy and processing parameters, should also shed further light into its underlying mechanisms. 

The impact of these results goes beyond fundamental considerations of nonlinear physics and oscillatory instabilities. Directional solidification of CMSX4 superalloy is of direct relevance to the production of single crystal turbine blades used in jet turbines. 
Therefore, the prediction of buoyancy-related defects and the stability of a CMSX4 solidification front is of immediate technological relevance to the casting of high-performance single-crystal components.

\section*{Acknowledgements}

This study was supported in part by the European Union’s Horizon 2020 research and innovation programme through a Marie Sk\l odowska-Curie Individual Fellowship (Grant Agreement 842795) and by the Spanish Ministry of Science through a Ramón y Cajal Fellowship (Ref. RYC2019-028233-I).
We, the authors, also wish to thank Guillaume Reinhart, Alexandre Viardin, and Ingo Steinbach, for providing data and/or figures necessary to compare our results to theirs.

\bibliography{bibliography_0.bib}

\pagebreak
\onecolumn
\begin{center}
\textbf{\large Supplemental Materials: {Convective effects on columnar dendritic solidification -- A multiscale dendritic needle network study }}\\[.5in]
\end{center}
\setcounter{equation}{0}
\setcounter{figure}{0}
\setcounter{table}{0}
\setcounter{page}{1}
\makeatletter
\renewcommand{\theequation}{S\arabic{equation}}
\renewcommand{\thepage}{S\arabic{page}}
\renewcommand{\thefigure}{S\arabic{figure}}

\appendix 
\renewcommand*{\thesection}{\Alph{section}}
~\\[-1.75cm]
\section{Kurz-Giovanola-Trivedi (KGT) model}
\label{sm_kgt}
The analytical KGT model for dendritic growth is based on the Ivantsov paraboloid solution and the marginal stability criterion and was introduced in Ref.~\cite{kurz1986} for binary alloys. It allows to predict the tip radius $R$ of a columnar dendrite, depending on the solidification conditions, namely temperature gradient $G$ and growth velocity $V$. The model was extended to ternary alloys \cite{bobadilla1988,rappaz1989,rappaz1990}, ignoring cross-species diffusion. 
This latter assumption essentially leads to simply adding up individual species contributions --- which is also consistent with the theroetical limits of planar instability for ternary alloys under a similar assumption \cite{coates1968solid}.
We use a generalization of this approach, briefly described below, to estimate the tip radius versus velocity of the multicomponent CMSX-4 alloy.

The solute supersaturation of each species $i$ is given by the Ivantsov solution \cite{ivantsov1947} 
\begin{align}
\label{eq:ivantsov}
\Omega_i = \text{Iv}(\text{Pe}_i)=\text{Pe}_i\exp(\text{Pe}_i)\text{E}_1(\text{Pe}_i)
\end{align}
with the solute supersaturation
\begin{align}
\label{eq:omega}
\Omega_i = (c_i-c_{\infty,i})/((1-k_i)c_i)
\end{align}
and the P\'eclet number
\begin{align}
\label{eq:peclet}
\text{Pe}_i=RV/(2D_i)~,
\end{align}
where $c_i$, $c_{\infty,i}$, $D_i$, and $k_i$ are, respectively, the concentration, the nominal concentration, the diffusion coefficient, and the partition coefficient of species $i$. 

In a $(N+1)-$component alloy, considering a linearized phase diagram and neglecting kinetic undercooling, the tip temperature $T$ of a growing dendrite is given by
\begin{align}
\label{eq:temp}
 T &= T_L + \sum^N_{i=1} \big\{ m_i(c_i-c_{\infty,i}) \big\} - \frac{2\Gamma}{R}
 = \underbrace{T_L-\sum^N_{i=1}\big\{m_ic_{\infty,i}\big\}}_{T^\prime_M}+\sum^N_{i=1}\big\{m_ic_i\big\} - \frac{2\Gamma}{R}
\end{align}
with the liquidus temperature $T_L$ and the Gibbs-Thomson coefficient $\Gamma$, and an artificial melting temperature $T^\prime_M$ of the pure solvent, extrapolated for the local slopes at $T_L$.

Marginal stability theory, similarly generalized to a multicomponent alloy, gives
\begin{equation}
 R=2\pi\sqrt{\frac{\Gamma}{\sum^N_{i=1}\big\{m_iG_{c,i}\xi(\text{Pe}_i,k_i)\big\}-G}}
\end{equation}
with
\begin{equation}
 \xi_c(\text{Pe}_i,k_i) = 1-\frac{2k_i}{\sqrt{1+\left(\frac{2\pi}{\text{Pe}_i}\right)^2}-1+2k_i}~,
\end{equation}
where $G_{c,i}$ is the solute gradient of species $i$ in the liquid ahead of the tip, given by
\begin{equation}
 G_{c,i}=-\frac{V}{D_i}c_i(1-k_i) = -\frac{V}{D_i}\frac{c_{\infty,i}(1-k_i)}{1-(1-k_i)\text{Iv}(\text{Pe}_i)}~.
\end{equation}

Combining the previous equations, one obtains the second order polynomial
\begin{equation}
\label{eq:kgt_0}
 \frac{4\pi^2\Gamma}{R^2}+\frac{2}{R}\sum^N_{i=1}\left\{\frac{\text{Pe}_im_ic_{\infty,i}(1-k_i)\xi_c(\text{Pe}_i,k_i)}{1-(1-k_i)\text{Iv}(\text{Pe}_i)} \right\}+ G = 0~.
\end{equation}
The dendrite tip temperature (Eq.~\eqref{eq:temp}) and radius (Eq.~\eqref{eq:kgt_0}) can thus be calculated iteratively.
Here, using a bespoke Python script, we set a velocity $V$, and then solve for $R$ and $T$ using Eqs \eqref{eq:kgt_0} and \eqref{eq:ivantsov}-\eqref{eq:temp} and iterating until convergence of the tip temperature between two iterations.

Table \ref{tab:parameters} contains all parameter used in our KGT calculations for both multicomponent and the binary surrogate alloy. The liquidus temperature $T_L$, liquidus slopes $m_i$ and partition coefficients $k_i$ are obtained via CalPhaD (ThermoCalc with TCNI8 database). For the solute diffusion coefficients $D_i$ of the respective species we use average values of the ones given in the Supplementary Material of Refs \cite{yang2019a} and \cite{leonard2004}.
The parameters for the binary alloy are obtained as discussed in the main article.

\begin{table*}[h]
\centering
\begin{tabular}{lrrrrrrrr}
\multicolumn{9}{c}{\textbf{Multicomponent CMSX-4}} \\
\hline
 & Cr & Co & W & Al & Mo & Re & Ti & Ta \\
\hline
 $c_{\infty,i}\,/\,\si{\wtpercent}$ & \num{6.5} & \num{9.6} & \num{6.4} & \num{5.6} & \num{0.6} & \num{3.0} & \num{1.0} & \num{6.5} \\
 $m\,\cdot\si{\wtpercent\per\kelvin}$ & \num{-595.3} & \num{-1.47} & \num{-302.9} & \num{-1308.6} & \num{-616.1} & \num{-28.4} & \num{-2025.5} & \num{-657.1} \\
 $k$ & \num{0.96} & \num{1.17} & \num{1.07} & \num{0.90} & \num{0.75} & \num{1.55} & \num{0.46} & \num{0.54} \\
 $D\,\cdot\SI{e9}{\second\per\meter^2}$ & \num{1.1} & \num{1.0} & \num{1.1} & \num{3.0} & \num{2.0} & \num{1.0} & \num{1.4} & \num{1.7} \\
\hline
\end{tabular}
\\[.2cm]
\begin{tabular}{lll}
\textbf{Binary surrogate alloy} &\quad& \textbf{Universal parameters} \\
\hline
 $c_{\infty}=\SI{6.6}{\wtpercent}$ && $T_M = \SI{1728}{\kelvin}$\\
 $m=\SI{-25}{\kelvin\per\wtpercent}$ && $T_L = \SI{1660}{\kelvin}$\\
 $k=\num{0.8}$ && $\Gamma = \SI{2.46e-7}{\kelvin\meter}$ \\
 $D=\SI{2e-9}{\meter^2\per\second}$ &&  $G = \SI{4400}{\kelvin\per\meter}$\\
\hline
\end{tabular}
\caption{Input parameters for the KGT model calculations for the CMSX-4 multicomponent alloy (ignoring hafnium), and the binary surrogate alloy. \label{tab:parameters}}
\end{table*}

\renewcommand*{\thesection}{\Alph{section}}
\section{Oscillatory growth in nickel-based CMSX-4 superalloy}

Here we provide more details and extended results from the two-dimensional DNN simulations of the surrogate alloy discussed in Sec.\,4 of the article. 

\begin{figure*}
\centering
\includegraphics[width=1\textwidth]{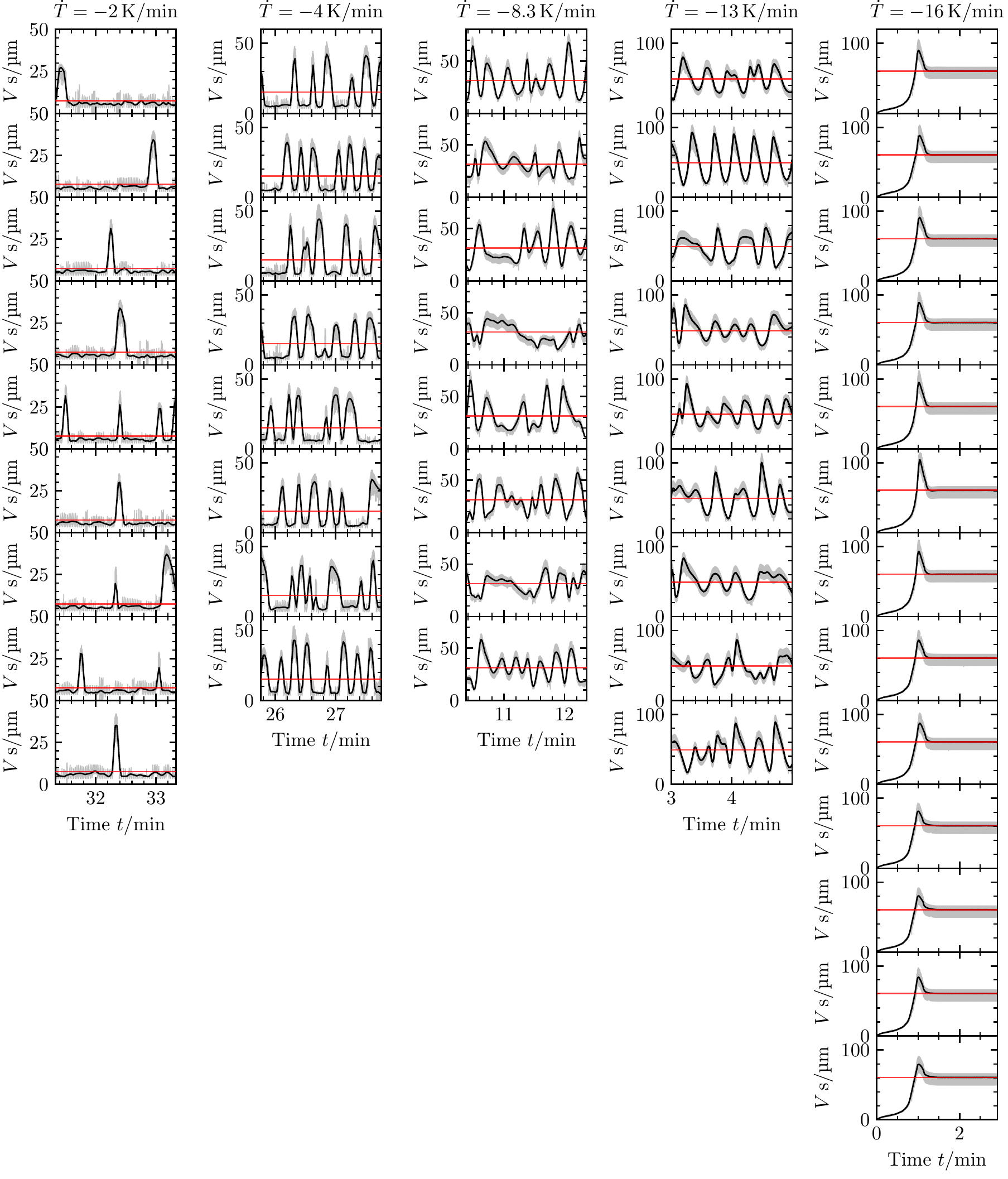}
\caption{Tip velocities predicted by DNN simulations at different cooling rates. Velocities of eliminated needles are not shown. The red lines represent the corresponding theoretical steady state growth velocity.\label{fig:nickel_based_velocities_R}}
\end{figure*}

Fig.\,\ref{fig:nickel_based_velocities_R} contains velocity plots of all non-eliminated needles of the DNN simulations at different cooling rates $\dot T$, corresponding to the ones shown in Fig.\,4 of the article. 

\begin{figure*}
\centering
\includegraphics[width=1\textwidth]{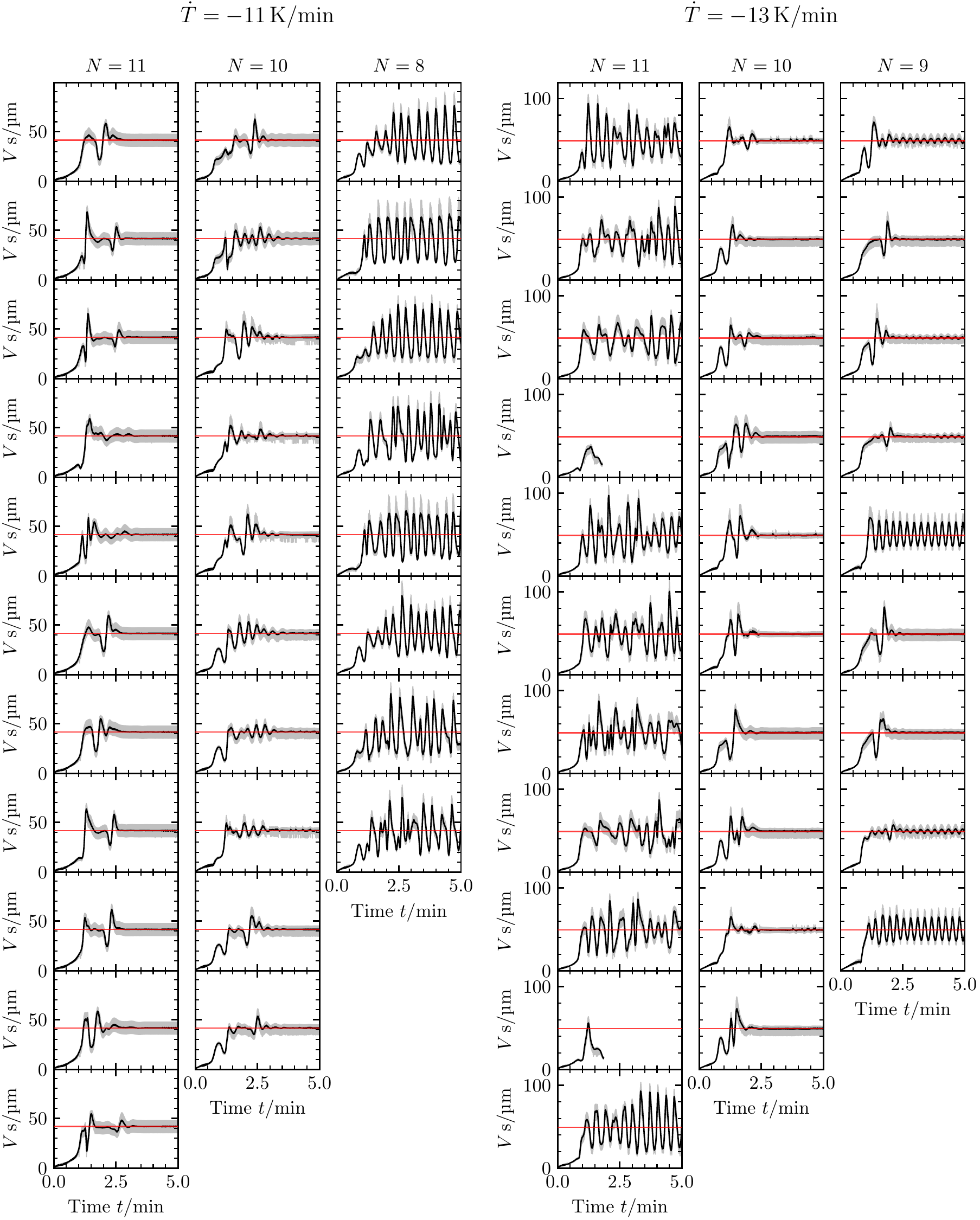}
\caption{Tip velocities predicted by DNN simulations with different initial number of needles $N$, at $\dot T=\SI{-11}{\kelvin\per\minute}$ and $\dot T=\SI{-13}{\kelvin\per\minute}$. The red lines represent the corresponding theoretical steady state growth velocity. \label{fig:nickel_based_velocities_N}}
\end{figure*}

Fig.\,\ref{fig:nickel_based_velocities_N} contains velocity plots of all needles of simulations with different initial number of needles $N$, at two different cooling rates. At $\dot T=\SI{-11}{\kelvin\per\minute}$, the oscillatory behavior is promoted by decreasing the number of initial needles, as discussed in the article. At $\dot T=\SI{-13}{\kelvin\per\minute}$ with $N=\num{11}$, the needle distribution becomes non-uniform after two needles (fourth from top and second from bottom) are eliminated. The corresponding simulation is illustrated in Fig.\,5 of the article. The large spacing favors plume formation, higher fluid velocities, and hence oscillations. The same simulation with $N=10$ results in damped fluctuations, since no large spacings are generated and the needle distribution stays uniform. For $N=9$, most needle velocities are damped, but oscillations start to emerge for some dendrites.

\begin{figure*}
\centering
\includegraphics[width=1\textwidth]{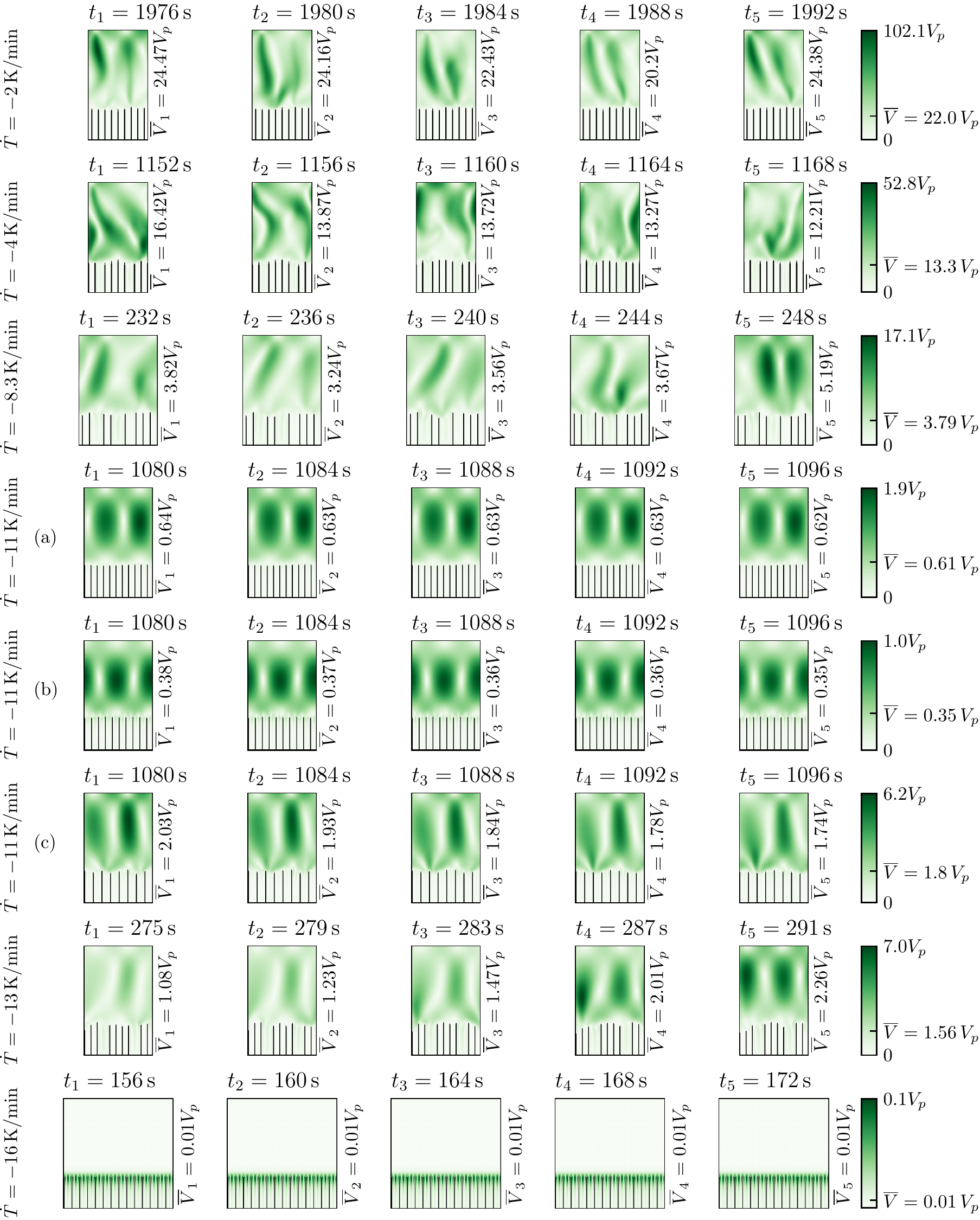}
\caption{Velocity magnitudes $V$ for five time snapshots $t_i$ at the late stage of the simulations, for each simulated cooling rate $\dot T$. The average velocities $\overline{V}$ are marked in the corresponding colorbars. \label{fig:velocit_average}}
\end{figure*}

Fig.\,\ref{fig:velocit_average} shows the map of the velocity magnitude for five different time steps for each simulation of Sec.~4, as well as the velocities $\overline V_n$ averaged over the entire liquid domain at these five time steps ($1\leq n\leq5$), which are the five values used to estimate the spatiotemporal averaged velocity $\overline V$ marked on the colorbar on the right-hand-side of the figure and listed in Table~6 of the article.

\begin{figure*}
\centering
\includegraphics[width=0.75\textwidth]{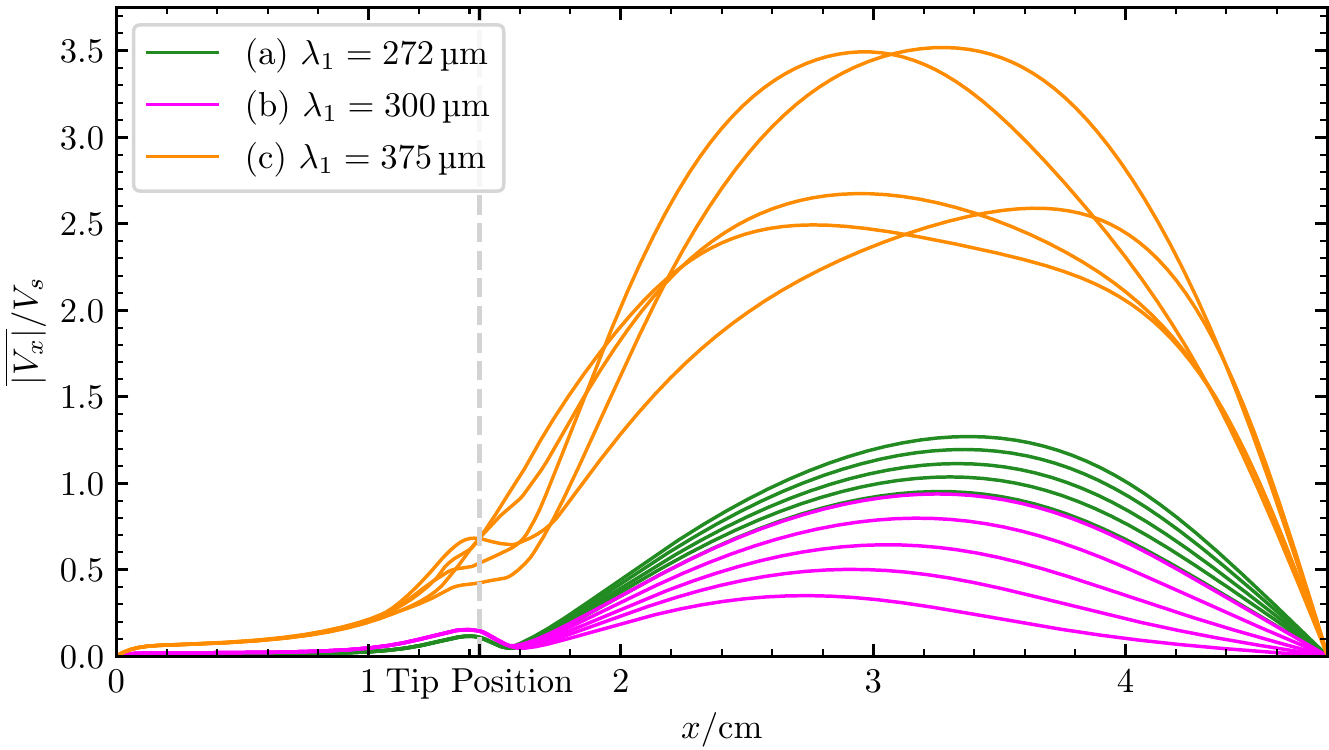}
\caption{Average magnitude of the $x$-component $\overline{|V_x|}$ of the melt velocity along the vertical $x$-direction  of the simulation domain for the cases (a)-(c) of Fig.\,6 in the main article at times $t = 246$, 259, 273, 286, and 300\,s.\label{fig:velocities_line_sample}}
\end{figure*}

\begin{figure*}
\centering
\includegraphics[width=0.75\textwidth]{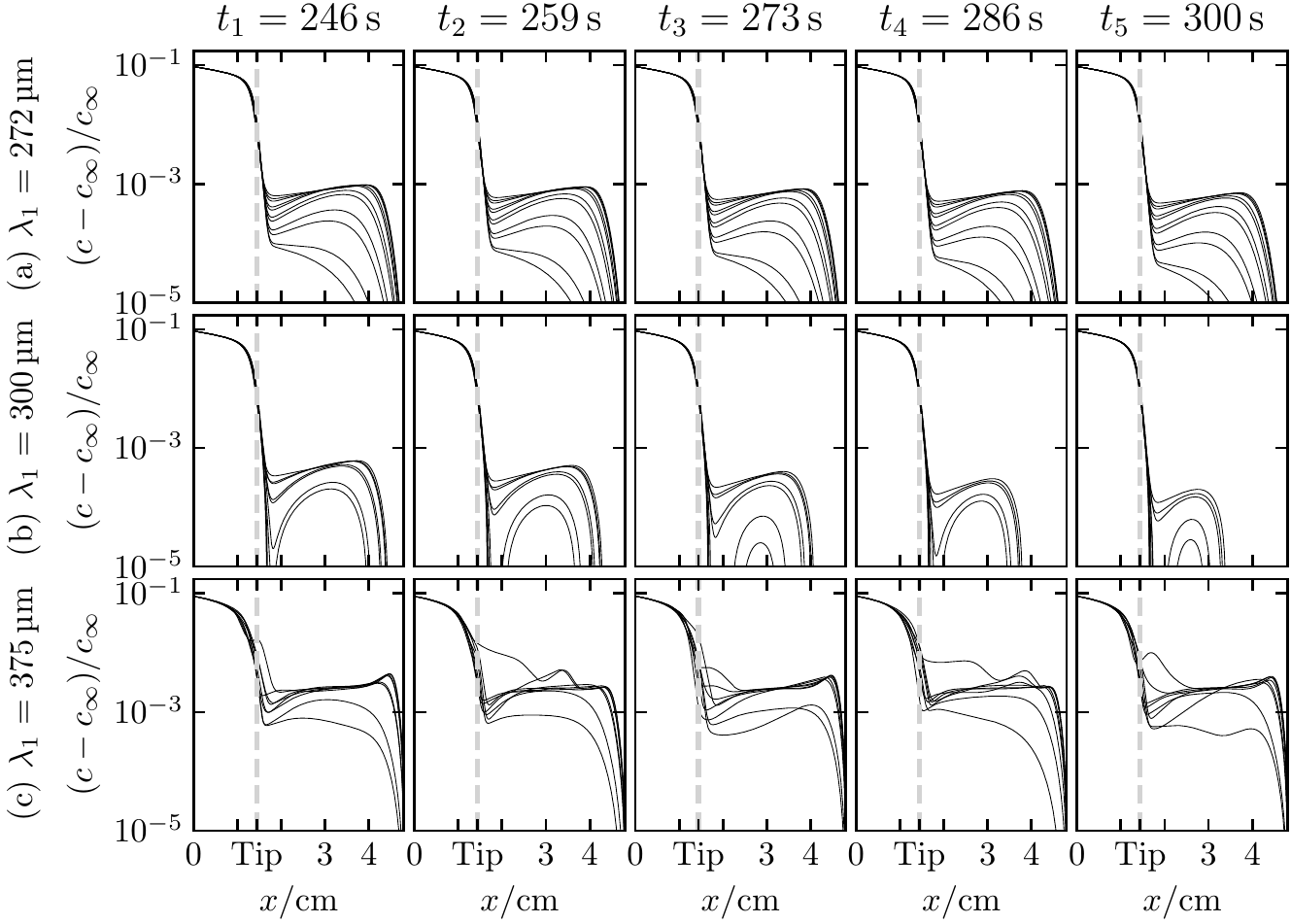}
\caption{Longitudinal concentration profiles for the cases (a)-(c) of Fig.\,6 in the main article, sampled along vertical lines parallel to the growth direction and located at the center between two adjacent needles, at different times (different columns).\label{fig:concentrations_line_sample}}
\end{figure*}

\begin{figure*}
\centering
\includegraphics[width=0.75\textwidth]{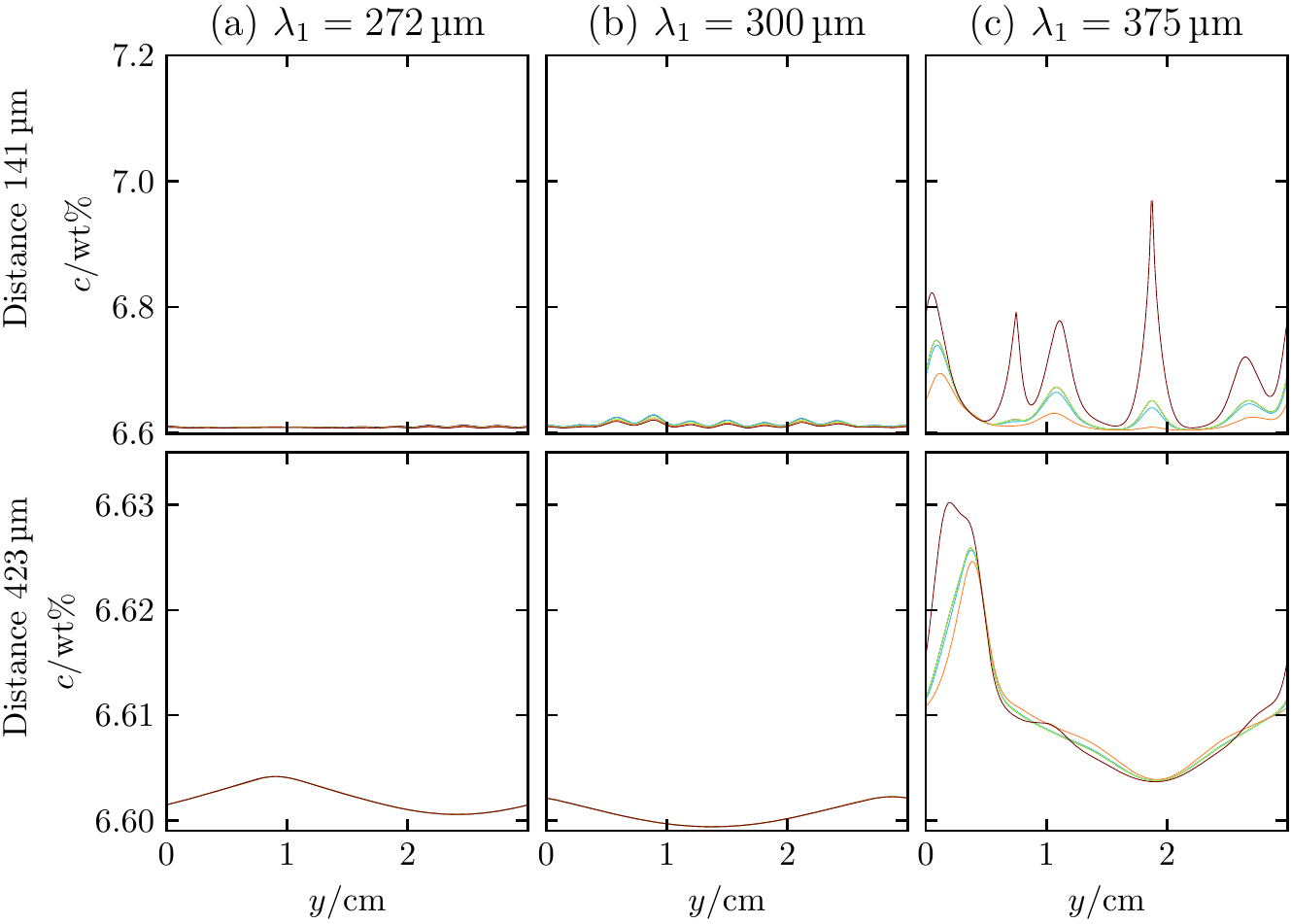}
\caption{Transversal concentration profiles, sampled along lines normal to the growth direction at $x=x_\text{tip}+\SI{141}{\micro\meter}$ and $\SI{423}{\micro\meter}$ ahead of the most advanced tip position ($x_\text{tip}$) at different times (different colored lines) for the cases (a)-(c) of Fig.\,6 in the main article.
\label{fig:concentrations_line_sample_horizontal}}
\end{figure*}

Fig.\,\ref{fig:velocities_line_sample} shows the vertical component of the velocity ($V_x$), averaged over the entire liquid region ($\overline{|V_x|}$) as a function of the height $y$ at five different times $t = 246$, 259, 536 273, 286, and 300\,s, which also correspond to times illustrated in figures \ref{fig:concentrations_line_sample} and \ref{fig:concentrations_line_sample_horizontal}.

Fig.\,\ref{fig:concentrations_line_sample} shows longitudinal composition profiles for the three cases of Fig,\,6 of the main article, sampled along lines parallel to the growth direction and located in the center between adjacent needles at different times.

Fig.\,\ref{fig:concentrations_line_sample_horizontal} shows transversal composition profiles for the three cases of Fig,\,6 of the main article, sampled along lines normal to the growth direction at $x=x_\text{tip}+\SI{141}{\micro\meter}$ and $\SI{423}{\micro\meter}$ ahead of the most advanced tip position ($x_\text{tip}$) at different times.

\end{document}